\documentclass[prb,twocolumn,floatfix,amsmath,amssymb,superscriptaddress,nobibnotes,aps,floatfix]{revtex4-2}
\usepackage{graphicx}% Include figure files
\usepackage{dcolumn}% Align table columns on decimal point
\usepackage{bm}% bold math
\usepackage{color}
\usepackage[dvipsnames]{xcolor}
\usepackage{hyperref}
\hypersetup{colorlinks=true,urlcolor=blue,citecolor=blue}
\usepackage{soul}
\usepackage{amsmath}

\begin{document}

\title{Skyrmion Phase Control by Magnetic Dipole-Dipole Interaction and Electric Field in Centrosymmetric Materials}
	
\author{Raz Rivlis}
\author{Alexander Petrovi\'c}
\author{Yuri Dahnovsky}
\affiliation{$^1$Department of Physics and Astronomy, University of Wyoming, Laramie, 82071 WY, USA.}
\affiliation{$^2$Center for Quantum Information Science and Engineering, University of Wyoming, Laramie, 82071 WY, USA.}
	
\date{\today}
	
\begin{abstract}
Establishing precise control over the helicity and spatial configuration of magnetic skyrmions will be essential to realize their promise in classical, analog and quantum computation applications. In this work, we explore the role of magnetic dipole-dipole interactions, external electric fields, and magnetic fields in controlling these parameters within a triangular lattice centrosymmetric skyrmion host. We demonstrate that dipole-dipole interactions strongly favor Bloch helicity. Notably, a zero magnetic field skyrmion phase appears upon raising the dipole-dipole coupling strength, with substantial potential for cost-effective quantum device applications. We also report the emergence of a meron/antimeron lattice phase, in the absence of any Dzyaloshinskii-Moriya interaction. In contrast, applied electric fields stabilize high density N\'eel skyrmion crystals. The interplay between dipole-dipole interactions and external electric fields creates a continuous transition between the two skyrmion types, rather than an abrupt switch. Applied electric fields can therefore be used as a continuous tuning mechanism for skyrmion helicity, and hence a control handle for tuning two-level systems in skyrmion qubits. 
\end{abstract}
	
%\pacs{Valid PACS appear here}
	
\maketitle
	
\section{Introduction}
	
In the past decade, noncollinear magnetic materials featuring topological spin textures have attracted considerable attention due to their potential technological applications~\cite{Nagaosa13,Bogdanov20,Tokura2021,Li23,Khatua23}. In particular, nanoscale particle-like spin textures called skyrmions are of great interest for novel memory devices~\cite{Marrows2021,Luo21,Vakili2021}, analog or neuromorphic processors~\cite{Prychynenko2018,Song2020a,Yokouchi2022} and even qubits~\cite{Psaroudaki2023}. Magnetic skyrmions were initially predicted in the 1980s \cite{Bogdanov89}, and experimentally verified more recently in the late 2000s~\cite{Muhlbauer09,Seki12}. These localized magnetic textures have integer topological charge, which grants them an energy barrier against annihilation~\cite{Wild2017} and hence an extended lifetime even outside the parameter space in which they are thermodynamically stable. This long-term stability is what makes them valuable for information processing in devices~\cite{Back20}. 

The difficulty in delivering skyrmion applications lies primarily in their control mechanisms. In particular, controlling the size, geometric configuration, and creation/annihilation of skyrmions are high priorities, yet challenging to achieve. Another crucial attribute is the microscopic spin configuration within the skyrmion, usually parametrized by its helicity, vorticity and core polarity~\cite{Gobel2021}. Skyrmions are often categorized into two classes, N\'eel and Bloch, with internal spin rotations analogous to N\'eel and Bloch magnetic domain walls \cite{Dong23}. However, intermediate ``hybrid'' skyrmions also exist~\cite{Verba2020,Kong2024}, as well as antiskyrmions which (for the same core polarity) have opposite vorticities~\cite{Nayak2017}. Existing methods of manipulating skyrmions typically use spin-polarized electric currents, which exert spin orbit or spin transfer torques to translationally reconfigure skyrmions~\cite{Litzius20} and can even reverse the helicity within certain specialized device geometries~\cite{Wu2024}.  Alternative parameters with the potential to modulate the skyrmion stability and helicity include external electric~\cite{Yao2020,Dai2023} and magnetic fields~\cite{Li2023g}, and temperature~\cite{Lv2022}. Although these could in principle be varied locally for more precise skyrmion control~\cite{Shustin23, Apostoloff24, Pyatakov15, Wang19, Wang22}, such methods have yet to be consistently implemented in functional skyrmion architectures and full-spectrum $2\pi$ helicity tunability remains an elusive target.
		
Skyrmions were initially hypothesized to exist (and originally discovered) in noncentrosymmetric materials or at heterointerfaces, where they are typically stabilized by the Dzyaloshinskii-Moriya interaction (DMI)~\cite{Roessler2006}. The DMI is an antisymmetric exchange interaction, which emerges in the absence of spatial inversion symmetry and can induce a variety of noncollinear spin textures, including helical, conical and skyrmion phases~\cite{Lancaster19,Morikawa13}. Their spin helicity and vorticity are usually fixed by the symmetry-controlled orientation of the DMI vector, with limited scope for texture control (especially for small skyrmions in materials with strong DMIs). However, skyrmions have recently been found to be stabilized by magnetic frustration in bulk centrosymmetric materials~\cite{Kawamura2025}. These materials exhibit no global DMI due to their crystal symmetry. Some examples include Gd$_2$PdSi$_3$~\cite{Gomil25, Takashi19, Hirschberger20, Zhang20}, GdGa$_2$~\cite{Baral25,Wang26}, EuPtSi~\cite{Matsumura2024} and EuNiGe$_3$~\cite{Matsumura2024a}, all of which have triangular lattices, and EuAl$_4$ which has a tetragonal crystal structure~\cite{Takagi22, Grant26, Shang21}.
		
Several physical mechanisms are, in principle, capable of stabilizing skyrmions in centrosymmetric systems, including short range geometric frustration~\cite{Okubo12,Leonov15,Lin2016b,Lohani19} as well as longer range electron-electron interactions, such as the Ruderman-Kittel-Kasuya-Yosida (RKKY) interaction~\cite{Bouaziz2022,Paddison22,Matsuyama23,Dong24,Arai2026} or Kondo coupling on a triangular lattice~\cite{Ozawa2017,Wang2020,Wang2023f}. These frustrated states are of particular interest because the energies of N\'eel, Bloch, and antiskyrmions are degenerate under ideal conditions (i.e., with no skyrmion-skyrmion interactions and no other applied fields). This degeneracy renders centrosymmetric skyrmion hosts excellent candidates for manipulating skyrmion phases and creating fine-tuned energy barriers between distinct skyrmion states. In combination with the smaller size of centrosymmetric skyrmions compared to DMI-stabilized skyrmions, this could allow the creation of high density memory cells or stable qubit arrays. Here, we distinguish between proposed topological and ``conventional'' qubits: skyrmion-based topological qubits are bilayer systems, in which one layer hosts skyrmions and the other is a superconducting material~\cite{Petrovic2025}. N\'eel helicity skyrmions may create vortices in the superconductor~\cite{Hals2016,Baumard2019} containing localized Majorana zero modes~\cite{Rex19}, whose fermionic occupancy upon fusion determines the qubit state~\cite{Nothhelfer22, Konakanchi23}. In contrast,  ``conventional'' skyrmion qubits exploit the quantized two-level systems accessible at milliKelvin temperatures in individual spin textures with radii below $\sim$~10\,nm, notably Bloch skyrmions with opposite helicities \cite{Psaroudaki21,Xia2023}. Successful qubit design hence mandates the creation of a system with tunable spin helicity (i.e. selectivity between Bloch or N\'eel skyrmions), where the state is controllable by external fields and temperature. 

In this work we apply Monte Carlo methods to computationally study magnetically frustrated skyrmion systems on a triangular lattice, similar to 2D Gd$_2$PdSi$_3$ or GdGa$_2$ \cite{Gomil25,Takashi19,Baral25,Wang26}. We primarily map out the phase boundaries, with particular attention to distinguishing skyrmion type. We examine two mechanisms that cause transitions between N\'eel and Bloch skyrmions, namely magnetic dipole-dipole interactions and applied electric fields. This work is structured as follows: first, we explain the details of the model employed. Second, we examine phase transitions dependent on exchange interactions (similar to RKKY), magnetic field, temperature, and anisotropy. Lastly, we introduce magnetic dipole-dipole interactions and electric fields, to demonstrate how they affect the skyrmion type and configuration. As a result of our computations, we present phase diagrams and simulated spin texture images of skyrmion liquid and skyrmion crystal phases, as well as helical stripe states. Additionally, we report a meron/antimeron phase stabilized by dipole-dipole interactions.
		
\section{Theory and Computational Details}
	
\begin{figure}[htbp]
	\includegraphics[clip=true, width=0.8\columnwidth]{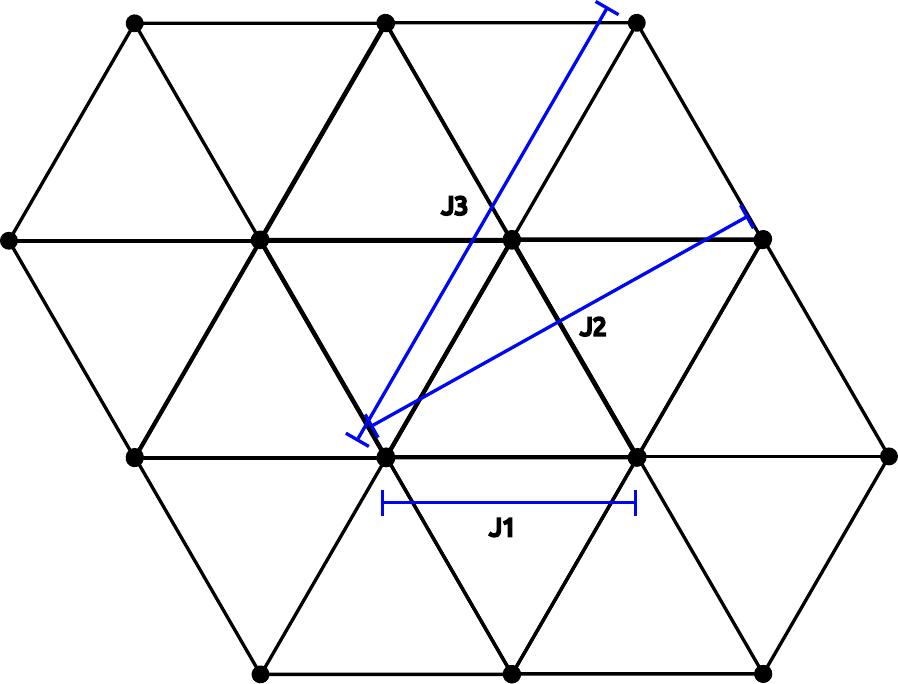}
	\caption{\label{schematic} Schematic illustrating the triangular crystal symmetry at the core of our theoretical model. $J_1$, $J_2$, and $J_3$ are the exchange interaction constants between first, second, and third nearest neighbor pairs, respectively.
	}
\end{figure}

The foundational model for all our simulations is a third nearest neighbor Heisenberg model on a triangular lattice. Figure~\ref{schematic} schematically demonstrates the lattice, showing examples of the pairs of the three nearest neighbors. The Hamiltonian for this model is:
\begin{widetext}
	\begin{equation}\label{heisenburg}		
		\begin{split}
			H = &-J_1 \sum_{\left< i, j \right>} \textbf{S}_i \cdot \textbf{S}_j - J_2 \sum_{\left<\left< i, j \right>\right>} \textbf{S}_i \cdot \textbf{S}_j - J_3 \sum_{\left<\left<\left< i, j \right>\right>\right>} \textbf{S}_i \cdot \textbf{S}_j - \sum_i \textbf{H} \cdot \textbf{S}_i - K \sum_i S_{i,z} ^2 .\\
		\end{split}
	\end{equation}
\end{widetext}
where $J_1$, $J_2$, and $J_3$ are the first, second, and third nearest neighbor interaction terms, $\textbf{S}$ is a classical spin vector, $\mathbf{H}$ is an applied magnetic field, and $K$ is an easy axis anisotropy term. The angled brackets under the first three sums denote sums over first, second, or third nearest neighbor terms. By choosing the proper signs and values of $J_2$ and $J_3$, we can model the RKKY interaction in centrosymmetric systems.
		
The defining characteristic of skyrmions is their integer topological charge, which can be calculated as follows~\cite{Heinze11}:
\begin{equation}
    Q = \frac{1}{4\pi}\int \textbf{S} \cdot \frac{\partial{\textbf{S}}}{ \partial{x}} \times \frac{\partial{\textbf{S}}}{ \partial{y}} dx dy.
\end{equation}
This is reduced to the sum of solid angles between spin vectors when applied to a discrete lattice. Topological charge is a measure of how many times the spin vectors wrap a sphere. All normal skyrmions have a topological charge of one, however, antiskyrmion structures are possible with a topological charge of $-1$. It is frequently convenient to distinguish two types of skyrmions, Bloch and N\'eel, which are named after the eponymous domain walls in ferromagnets and differ by their helicity $\varphi_0$. The skyrmion helicity describes the difference in angle between the azimuthal components of the skyrmion magnetic moments and the radial vector of the skyrmion~\cite{Nagaosa13,Gobel2021}. 

\begin{figure}[htbp]
	\includegraphics[clip=true, width=0.95\columnwidth]{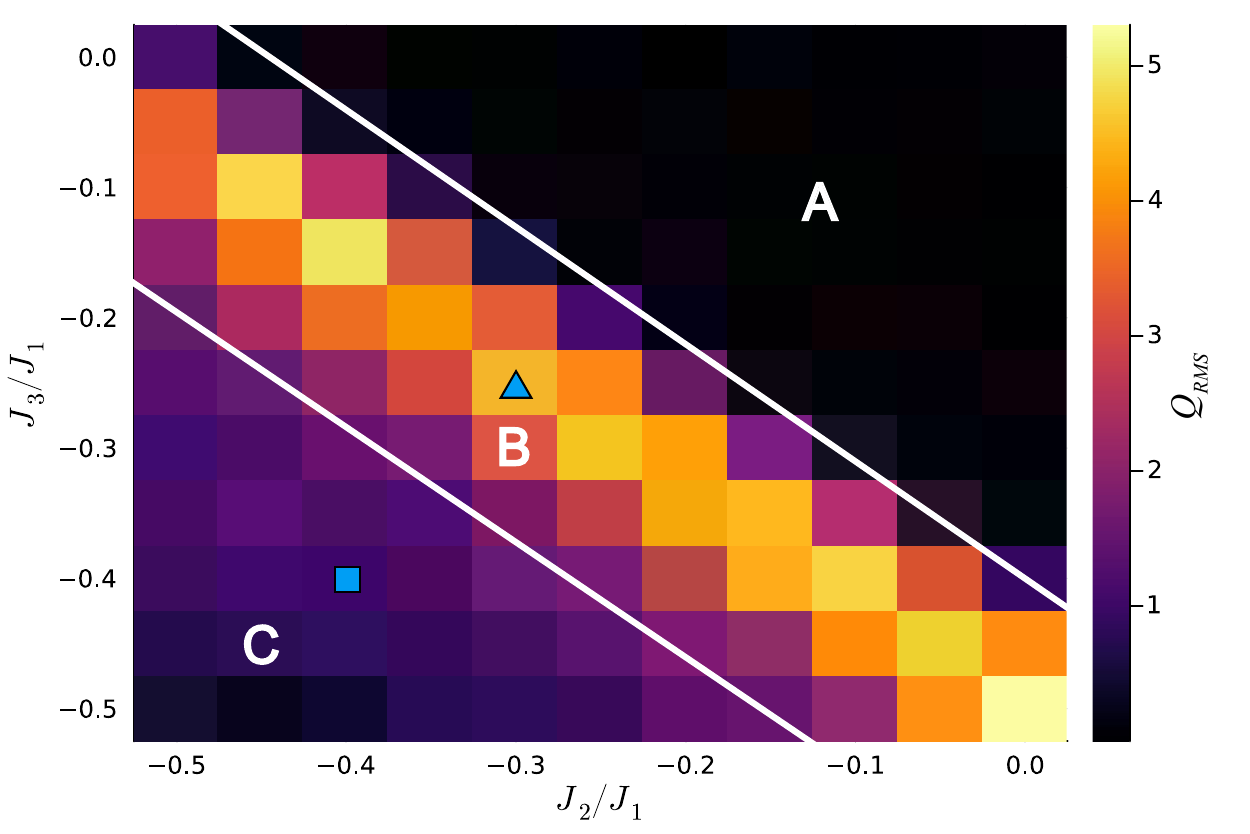}
	\caption{\label{j2j3phase} Phase diagram illustrating the magnetic phase evolution as a function of the second and third nearest neighbor exchange interactions. The plot is calculated at $B_z / J_1 = 0.3$ in the zero temperature limit. Phase A is a ferromagnetic phase. Phase B is a high skyrmion density phase, with mixed skyrmion types. Phase C is a helical phase, which features metastable skyrmions close to the boundary with phase B. Due to the presence of strong fluctuations between neighboring phases, all phase boundaries (white lines) in this work are qualitatively defined and serve as guides to the eye. The blue triangle and square denote the coordinates at which the spin textures shown in Fig.~\ref{j2j3images}(a) and (b) are evaluated, respectively. 
	}
\end{figure}
    
To study phase transitions, we employ the Metropolis-Hastings algorithm to simulate temperature-dependent magnetic phases with periodic boundary conditions~\cite{Hastings70}. Initial conditions are fully random, i.e., infinite temperature. Temperatures are then stepped down to allow some annealing, while still permitting metastable skyrmions and other noncollinear states to be found. Temperatures are determined by an arbitrary scale factor, and can be rescaled at will. We have chosen a realistic energy scale of $J_1 = 0.02$\,eV. Phase diagram plots are taken by averaging calculated values over 30 individual simulations to create a statistical ensemble. A measurement of the root mean square (RMS) value of the topological charge is used to find mixed skyrmion phases in which the net topological charge is zero, as well as measure the ratio of skyrmions to antiskyrmions. The RMS value is defined as:
\begin{equation}
	Q_{RMS} = \sqrt{\sum_n Q_n^2},
\end{equation}
where the sum is over the 30 individual computations performed for each chosen set of parameters with random initial conditions. Over large ensembles, the average value of $Q$ will be zero in a mixed skyrmion/antiskyrmion state; however, individual simulations may randomly display more of one type of skyrmion than another. Regardless of such variations, our RMS measurement of $Q$ allows us to determine phases and states containing any skyrmions in the most accurate way.
\section{Results}
\subsection{$J_{2}-J_{3}$ interactions}
To establish a starting point for our magnetically frustrated model, in Fig.~\ref{j2j3phase} we present a plot of the RMS  topological charge at zero temperature and zero anisotropy, demonstrating the dependence on the frustration terms $J_2/J_1$ and $J_3/J_1$. We calculate the RMS topological charge because mixed skyrmions and antiskyrmion phases would not be detected in the average topological charge, as the opposite topological charges cancel over large sample sizes. This plot is used to find different phases within our model. 

\begin{figure}[bp]
    \includegraphics[width=0.95\columnwidth]{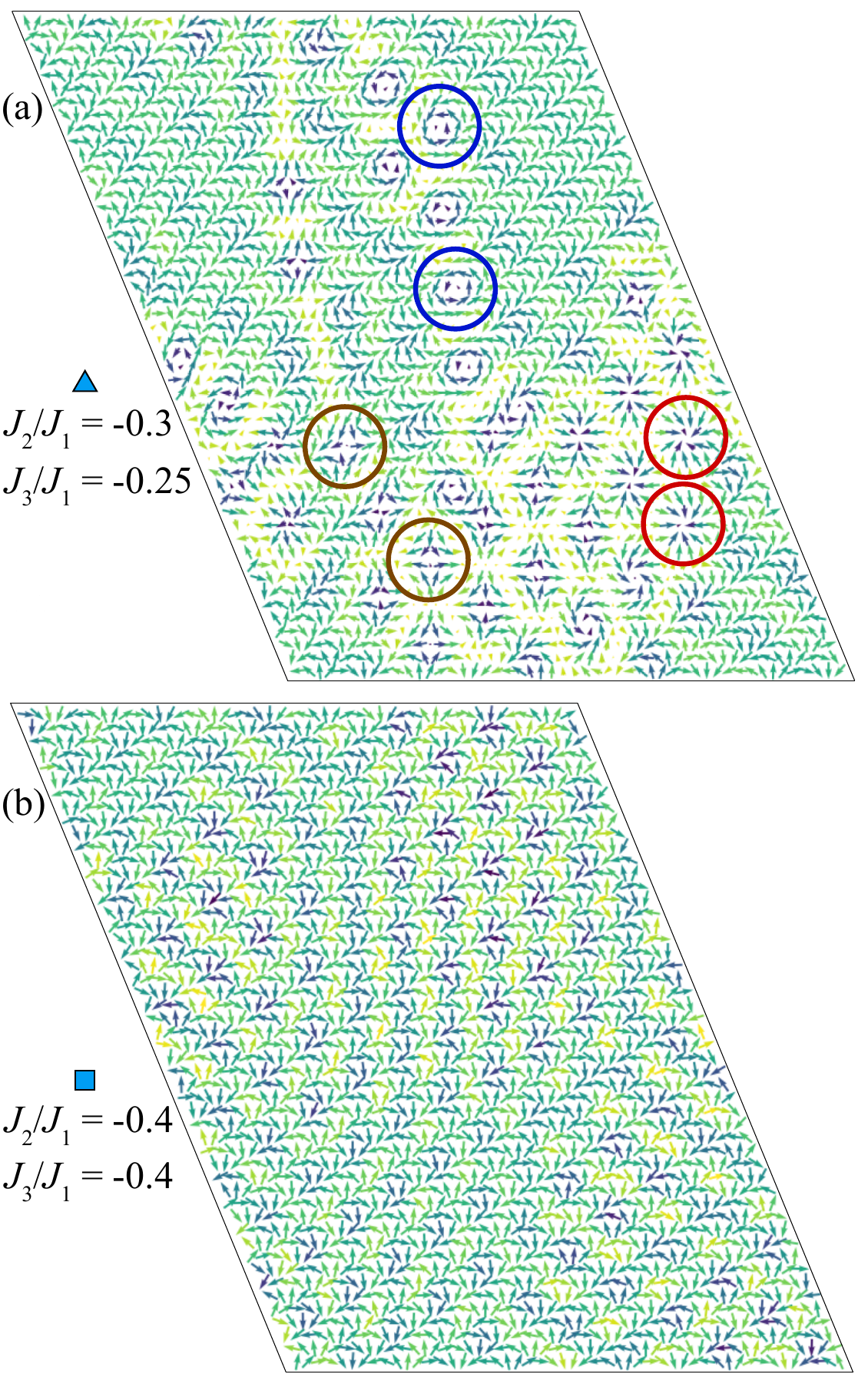}
	\caption{\label{j2j3images} Real-space spin textures evaluated at zero temperature, from the phase diagram shown in Fig. \ref{j2j3phase}. (a) Phase B (high-density skyrmions), illustrating the mixture of different skyrmion helicities and topological charges. The blue circles highlight several Bloch skyrmions, the red circles show N\'eel skyrmions, and the brown circles indicate antiskyrmions. (Not all skyrmions are highlighted, to preserve clarity). (b) Phase C (helical phase). $B_z/J_1$ is fixed at 0.3 for both phases.
    }
\end{figure}

In Fig.~\ref{j2j3phase}, the black sections correspond to zero topological charge, indicating a ferromagnetic phase, or a pure spin spiral phase, which are both topologically trivial (phases A and C, respectively). Phase B has the highest RMS value of the topological charge, indicating the presence of skyrmions, but not determining their type. Example images of individual simulations representative of the phases found in Fig.~\ref{j2j3phase} are shown in Fig.~\ref{j2j3images}. Of particular note is that in Fig.~\ref{j2j3images}(a), we find a mixed skyrmion phase, including N\'eel skyrmions, Bloch skyrmions, and antiskyrmions in a background of non-topological spin spirals. This indicates that all skyrmions in this phase are energy degenerate. Figure~\ref{j2j3images}(b) depicts the spin spiral state found in phase C from Fig.~\ref{j2j3phase}. 

\begin{figure}[htbp]
	\includegraphics[clip=true, width=0.95\columnwidth]{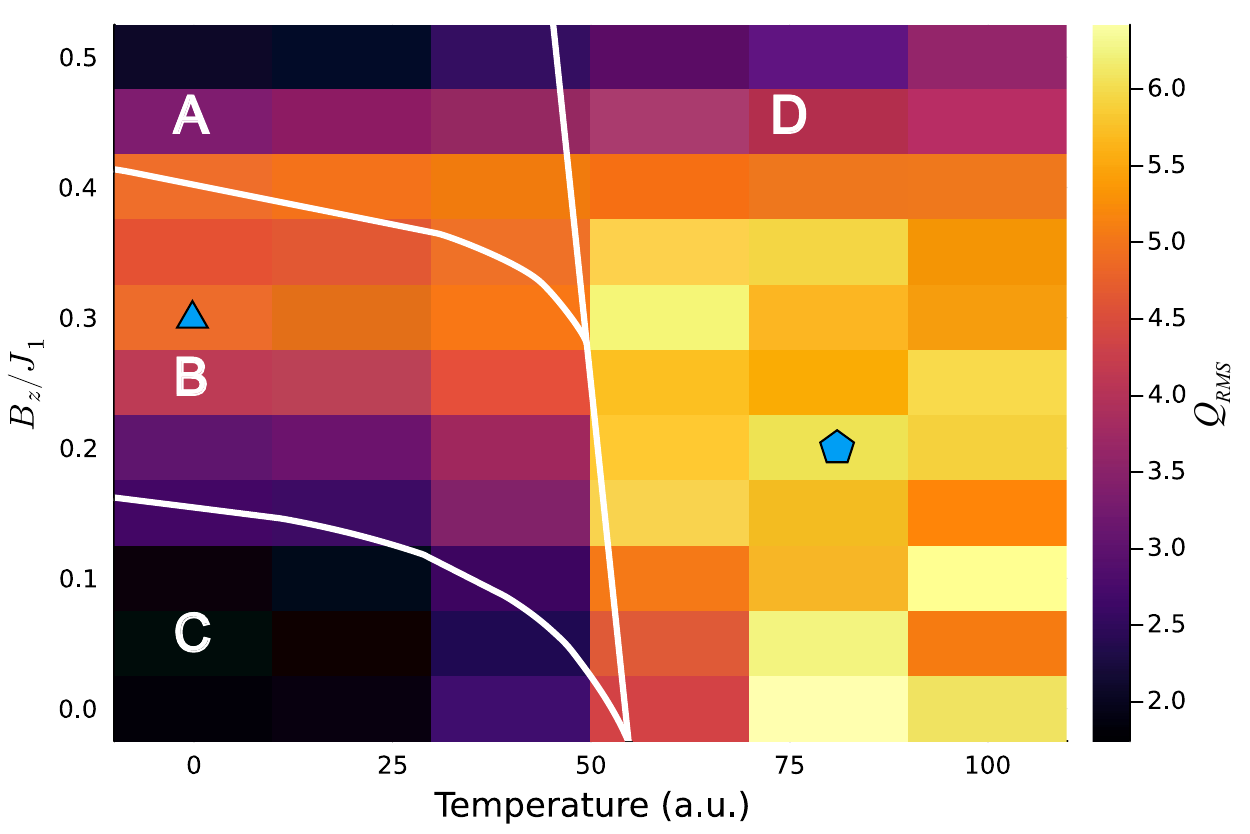}
	\caption{\label{tempbzphase} Phase diagram illustrating the magnetic phase distribution as a function of temperature and perpendicular magnetic field. The plot is calculated at $J_2 / J_1 = -0.3$ and $J_3 / J_1 = -0.25$. Phase A is a ferromagnetic phase. Phase B is a high skyrmion density phase, with mixed skyrmion types. Phase C is a helical phase exhibiting metastable skyrmions close to the boundary with phase B. Phase D is a high temperature paramagnetic phase; its high topological charge originates from short timescale chiral spin fluctuations. The appearance of the high density skyrmion phase B is in good agreement with the phase diagram in Ref.~\cite{Gomil25}. The magnetic ordering in phases A and C is qualitatively different from the phases observed in Gd$_2$PdSi$_3$, since this preliminary diagram does not include dipole-dipole interactions. The blue triangle and pentagon denote the coordinates at which the spin textures shown in Fig.~\ref{j2j3images}(a) and Fig.~\ref{HighTtexture} are evaluated, respectively.
	}
\end{figure}

The values of $J_2/J_1$ and $J_3/J_1$ are material-dependent properties. Therefore, specific values of $J_2/J_1$ and $J_3/J_1$ are chosen to match experimental results in known materials. Fig.~\ref{tempbzphase} shows a temperature-magnetic field phase diagram for $J_2/J_1 = -0.3$ and $J_3/J_1 = -0.25$. Phases A, B, and C in this plot are the same as phases A, B, C in Fig.~\ref{j2j3phase}, explored in a different parameter space. The emergence of the high density skyrmion phase B at intermediate fields is well aligned with ac susceptibility measurements for Gd$_2$PdSi$_3$ performed by Gomilsek et. al.~\cite{Gomil25}. These chosen values of $J_2/J_1$ and $J_3/J_1$ are used throughout the rest of this work to investigate the role of other interactions applied to a realistic material. 
		
\begin{figure}[htbp]
	\includegraphics[clip=true, width=0.95\columnwidth]{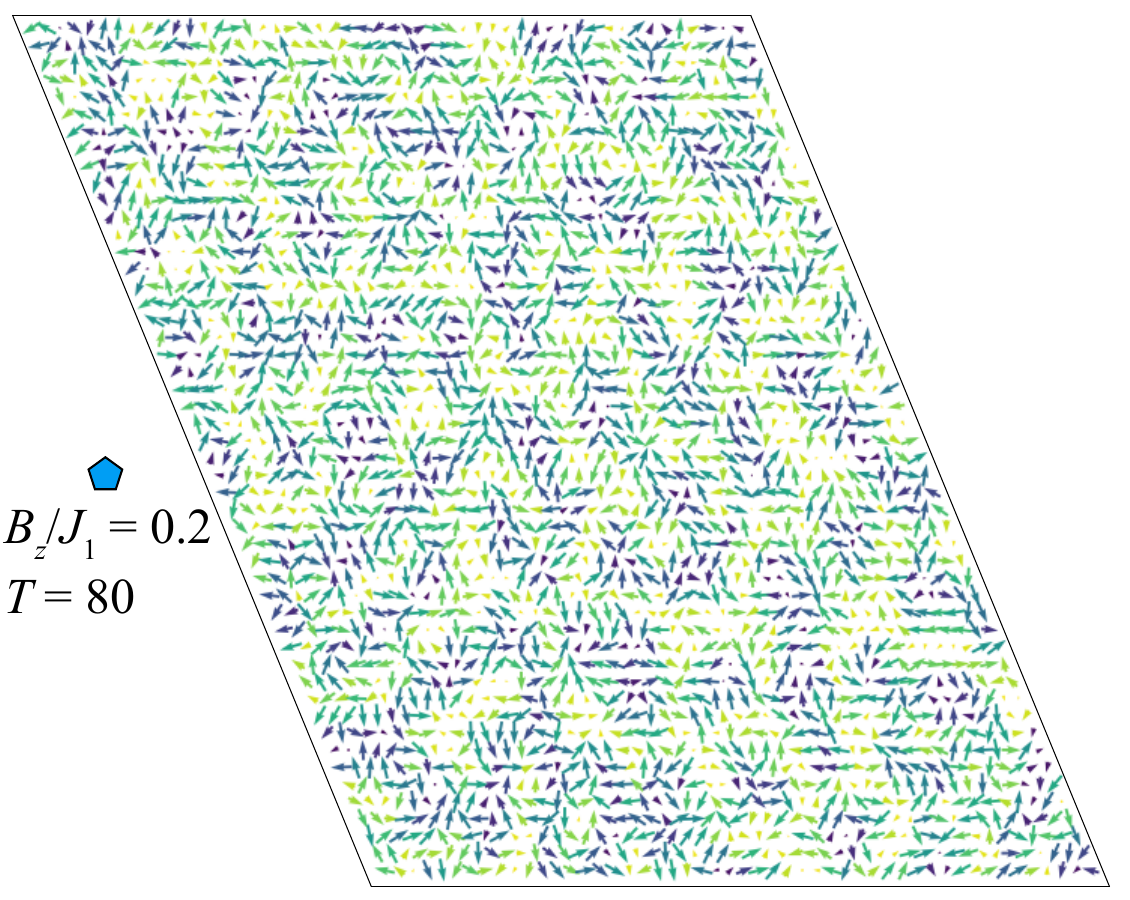}
	\caption{\label{HighTtexture} Spin configuration in the high temperature paramagnetic phase (blue pentagon in Fig.~\ref{tempbzphase}), evaluated at $J_2/J_1 = -0.3$, $J_3/J_1 = -0.25$.
	}
\end{figure}

Figure~\ref{HighTtexture} shows a snapshot of phase D found in Fig.~\ref{tempbzphase}. This phase is a high temperature paramagnetic phase. The topological charge is maximal in this phase due to short timescale chiral fluctuations~\cite{Fujishiro21,Raju2021}. 
	
\begin{figure}[htbp]
	\includegraphics[clip=true, width=0.95\columnwidth]{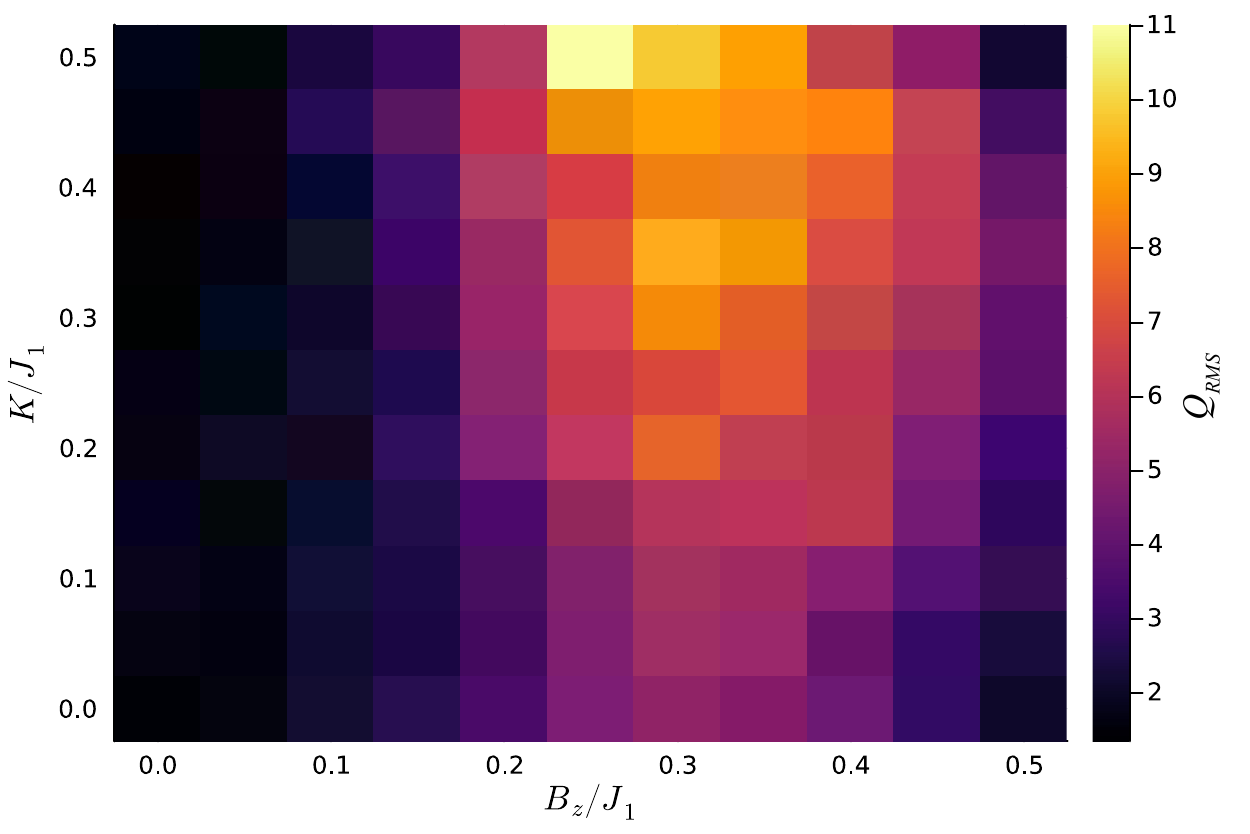}
	\caption{\label{bzani}  RMS values of the topological charge as a function of magnetic field and easy axis anisotropy. The phase diagram is calculated at $J_2/J_1 = -0.3$ and $J_3/J_1 = -0.25$ in the zero temperature limit.
	}
\end{figure}

In Fig.~\ref{bzani}, we present a $Q_{RMS}(B_z,K)$ phase diagram to show how including an easy axis anisotropy can help to stabilize skyrmions in a perpendicular magnetic field $B_z = \mu_0 H_z$. At low anisotropy, the skyrmion phase is found in a small range of magnetic fields ($B_z/J_1$ ranging from 0.2 to 0.4). With a higher anisotropy, skyrmions exist in a larger range of magnetic fields, and their concentration increases.	
	
\subsection{Magnetic dipole-dipole interaction}
	
\begin{figure*}[htbp]
   	\includegraphics[clip=true, width=1.9\columnwidth]{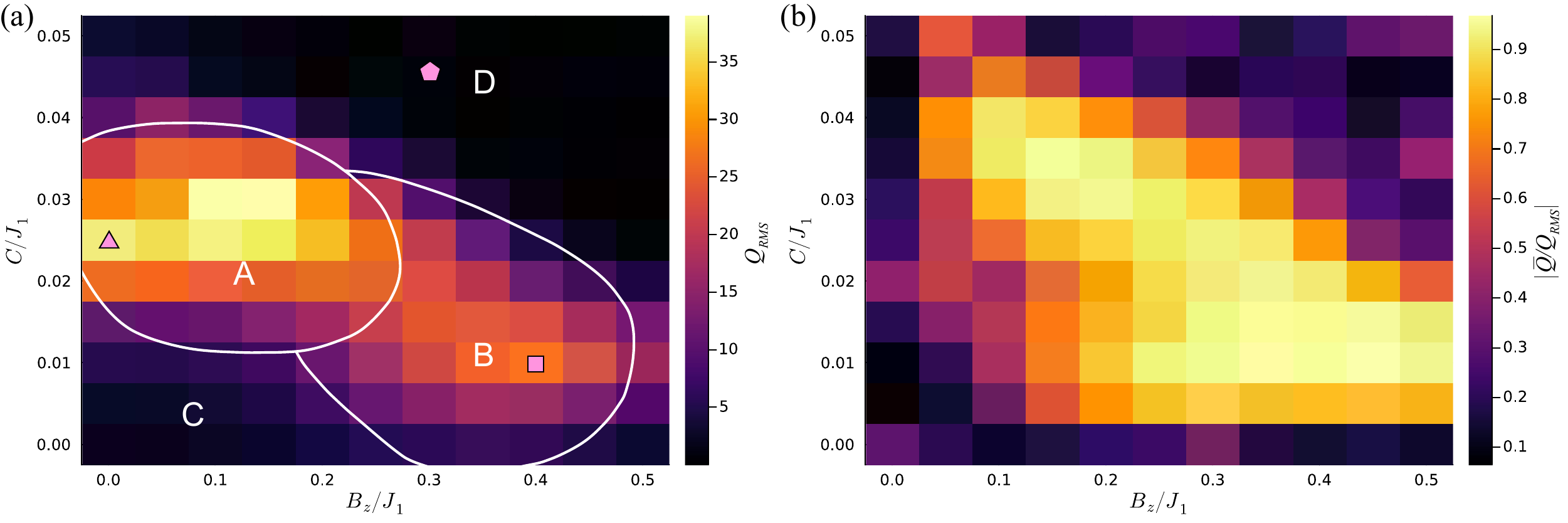}
	\caption{\label{bzdipolephase} Influence of perpendicular magnetic fields versus magnetic dipole-dipole interactions on the magnetic ordering. (a) Zero-temperature phase diagram, evaluated at $J_2 = -0.3$ and $J_3 = -0.25$, with the anisotropy $K$ fixed at zero. Phase A is a dense skyrmion crystal phase; phase B is a lower density skyrmion liquid. Phase C is a spin spiral state and phase D is a meron/antimeron crystal. Examples of each phase at the points indicated by the pink triangle, square and pentagon are shown in Fig.~\ref{bzdipoleimages}(a,b,c), respectively. (b) The same region of phase space as shown in (a), characterized by the ratio $\bar{Q}/Q_{RMS}$ to illustrate the balance between skyrmions and antiskyrmions. 
    }
\end{figure*}

The interactions between magnetic moments considered so far have been restricted to Heisenberg interactions, as detailed in Eq. \ref{heisenburg}. However, there is another magnetic interaction that must be taken into account: the magnetic dipole-dipole interaction. This interaction is described by the following Hamiltonian term: \cite{Jackson99, Jefremovas25} 
\begin{equation} \label{dipole}
	H = -\frac{\mu_0}{4\pi|\textbf{r}|^3} \sum_{i,j}[3 (\textbf{m}_i\cdot \hat{\textbf{e}}_{i,j})(\textbf{m}_j\cdot \hat{\textbf{e}}_{i,j}) - \textbf{m}_i \cdot \textbf{m}_j],
\end{equation}
where $\textbf{m}$ is the total magnetic moment, $|\textbf{r}|$ is the distance between the pairs of magnetic moments, and $\hat{\textbf{e}}$ is the unit vector in the direction between the magnetic moment pairs. In reduced units, Eq. \ref{dipole} can be rewritten as follows:
\begin{equation}
	H = -\frac{C}{|\textbf{r}|^3} \sum_{i,j}[3 (\textbf{S}_i\cdot \hat{\textbf{e}}_{i,j})(\textbf{S}_j\cdot \hat{\textbf{e}}_{i,j}) - \textbf{S}_i \cdot \textbf{S}_j],
\end{equation}
where $C$ is a constant in the units of $J_1$ and $\textbf{r}$ is measured in units of lattice spacings. $C$ represents the strength of the dipole-dipole interaction within a material, and is dependent on its lattice parameter.
    
It is important to understand how dipole-dipole interactions - which are ubiquitous in real magnetic materials - affect the skyrmion phases. Figure~\ref{bzdipolephase}(a) presents a phase diagram in which magnetic dipole-dipole interactions are progressively introduced, where the RMS topological charge $Q_{RMS}$ is plotted as a function of dipole-dipole interaction strength $C$ and perpendicular magnetic field. We clearly see four phases, where phases A and B are skyrmion crystal and liquid phases, respectively. When the dipole-dipole interaction is zero, $Q_{RMS}$ is small or zero. As the dipole-dipole interaction strength is increased, $Q_{RMS}$ rises quickly. This indicates two possibilities: either the skyrmion density is increasing, or the concentration of skyrmions vs. antiskyrmions is increasing, i.e.  the antiskyrmion density is falling. Phase C is the same spin spiral phase found in Figs.~\ref{j2j3phase} and~\ref{tempbzphase}. 
	
\begin{figure}[t]
	\includegraphics[width=0.95\columnwidth]{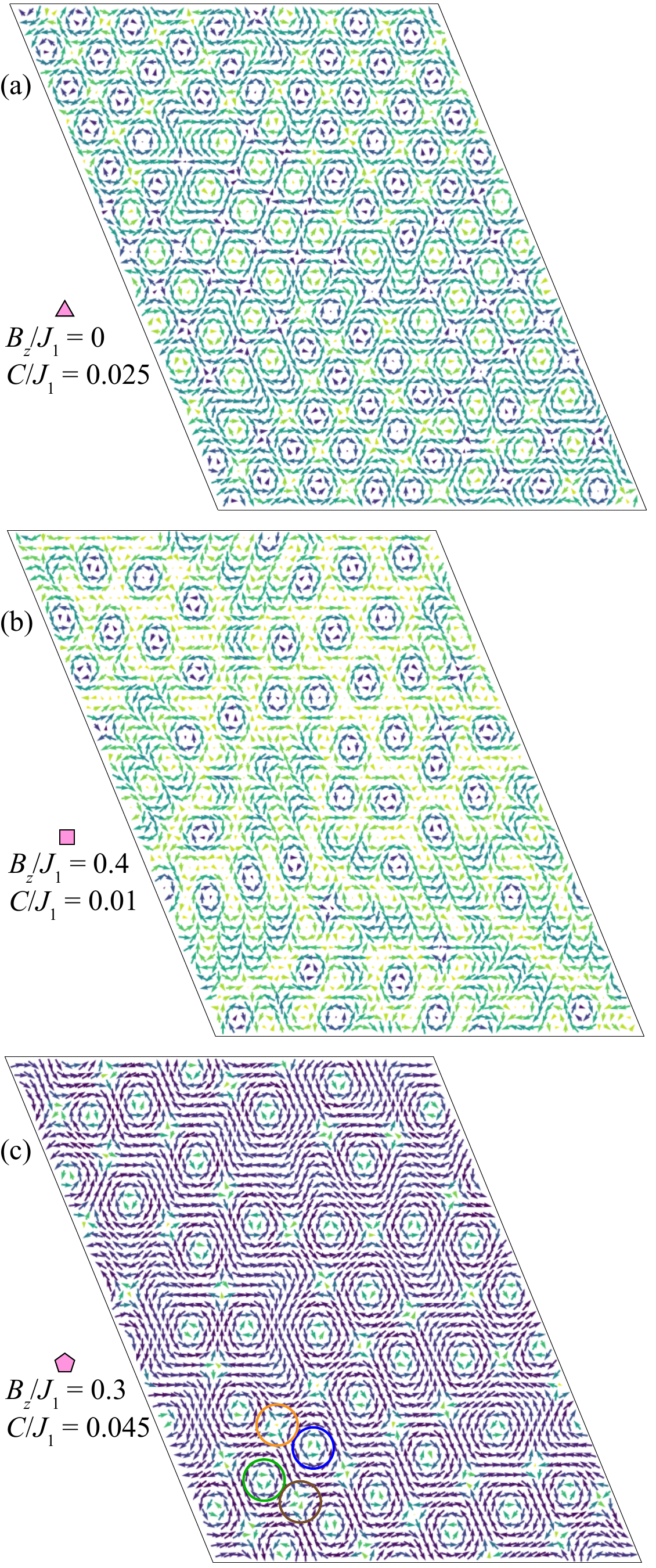}
\end{figure}
\begin{figure}[t]
	\caption{\label{bzdipoleimages} Spin textures stabilized upon varying the perpendicular magnetic field and dipole-dipole interaction as indicated in Fig.~\ref{bzdipolephase}. (a) Phase A is a skyrmion crystal, with Bloch helicity due to the strong dipole-dipole coupling. Since this simulation is performed in zero magnetic field, there is no preferred spin polarity (i.e., $\pm z$ spin orientation) at the center of each skyrmion.  Note the presence of lattice defects in the crystal, due to the intentionally limited annealing performed in our simulations. (b) Phase B is a dilute liquid of Bloch skyrmions.  (c) Phase D is a meron/antimeron crystal. Blue and green circles denote neighboring merons with Bloch helicities $\varphi_0=\pm\pi/2$, while brown and orange circles indicate antimerons rotated by $\pi/2$ with respect to each other. $J_2/J_1 = -0.3$, $J_3/J_1 = -0.25$ and $T=0$ for all three phases.
	}
\end{figure}

Comparatively, the skyrmion crystal phase A emerges at much lower fields than without dipole-dipole interactions. Notably, a finite skyrmion density is found even at zero magnetic field. This is very important for practical devices, as the skyrmions do not require a high field environment to remain thermodynamically stable. A snapshot of this phase is shown in Fig.~\ref{bzdipoleimages}(a). Phase B is related to the skyrmion phases found in Fig.~\ref{j2j3phase}, but with a much higher value of $Q_{RMS}$: a snapshot is shown in Fig.~\ref{bzdipoleimages}(b). From these snapshots, we observe far fewer antiskyrmions, and all skyrmions are Bloch type skyrmions. Additionally, a higher concentration of Bloch skyrmions is found in Fig.~\ref{bzdipoleimages}(a). Even fewer antiskyrmions are found in this state than in the phase represented by Fig.~\ref{bzdipoleimages}(b). This indicates that dipole-dipole interactions both increase the number of Bloch skyrmions and remove antiskyrmions. The most unexpected result is a new phase D [Fig.~\ref{bzdipoleimages}(c)], which only emerges upon the introduction of dipole-dipole interactions, and can be described as a meron/antimeron phase. Merons have a topological charge of $1/2$, but the phase has no net topological charge as there is a lattice of merons and antimerons with equal concentrations, canceling the total topological charge. Other meron/antimeron phases have previously been reported in DMI systems~\cite{Yu18,Hayami21,Mohylna25}. 
	
Besides using spin texture snapshots to determine the phase characteristics, we plot the average topological charge $\bar{Q}$ divided by $Q_{RMS}$ in Fig.~\ref{bzdipolephase}(b). This calculation provides a measure of the concentration of skyrmions vs. the concentration of antiskyrmions. When the concentrations are equal, the average value of $Q$ is zero, but when only skyrmions are present, $\bar{Q}=Q_{RMS}$. Any deviation between these two quantities hence indicates a mixed phase of skyrmions and antiskyrmions. As the dipole-dipole interaction strength, $C$, is increased, $\bar{Q}/Q_{RMS}$ quickly rises, indicating a far greater population of regular skyrmions compared to antiskyrmions. In fact, the removal of antiskyrmions is almost complete with even a small value of the dipole-dipole interaction magnitude. This topological charge selectivity is reinforced by spin texture images of the phases in Fig.~\ref{bzdipolephase}, which are presented in Fig.~\ref{bzdipoleimages}(a-c). These diagrams also display a helicity selectivity, with Bloch skyrmions ($\varphi_0=\pm\pi/2$) exclusively formed. We thus conclude that dipole-dipole interactions will efficiently remove antiskyrmions and N\'eel skyrmions from a centrosymmetric magnet, even with small strengths $<0.01\,J_1$. 
	
\subsection{Electric field}
The magnetoelectric effect is a known phenomenon where electric polarization can be induced by magnetic order, and strongly depends on magnetic symmetry groups~\cite{BLY83}. In certain materials with inhomogeneous magnetic fields (e.g., skyrmions, merons, domain walls, etc.), the magnetoelectric effect occurs as described in Refs.~\cite{Katsura05, Nikolaev19}. According to Katsura \textit{et al.}, the electric polarization vector is presented by the following equation~\cite{Katsura05}:
\begin{equation}
	\textbf{P}= P_0 \sum_{\left<i,j\right>} \hat{\textbf{e}}_{i,j} \times \textbf{S}_i \times \textbf{S}_j,
\end{equation}
where $P_0$ is a material-dependent parameter. Here we estimate $P_0$ to be approximately $10^{-3}$ C/m$^2$, corresponding to physically plausible electric field strengths of approximately $10^3$ V/cm~\cite{Nikolaev19}. Thus, the additional Hamiltonian term introduced by the magnetoelectric effect is: 
\begin{equation}
	H = -\textbf{E} \cdot \textbf{P}.
\end{equation}

\begin{figure*}[htbp]
	\includegraphics[clip=true, width=1.9\columnwidth]{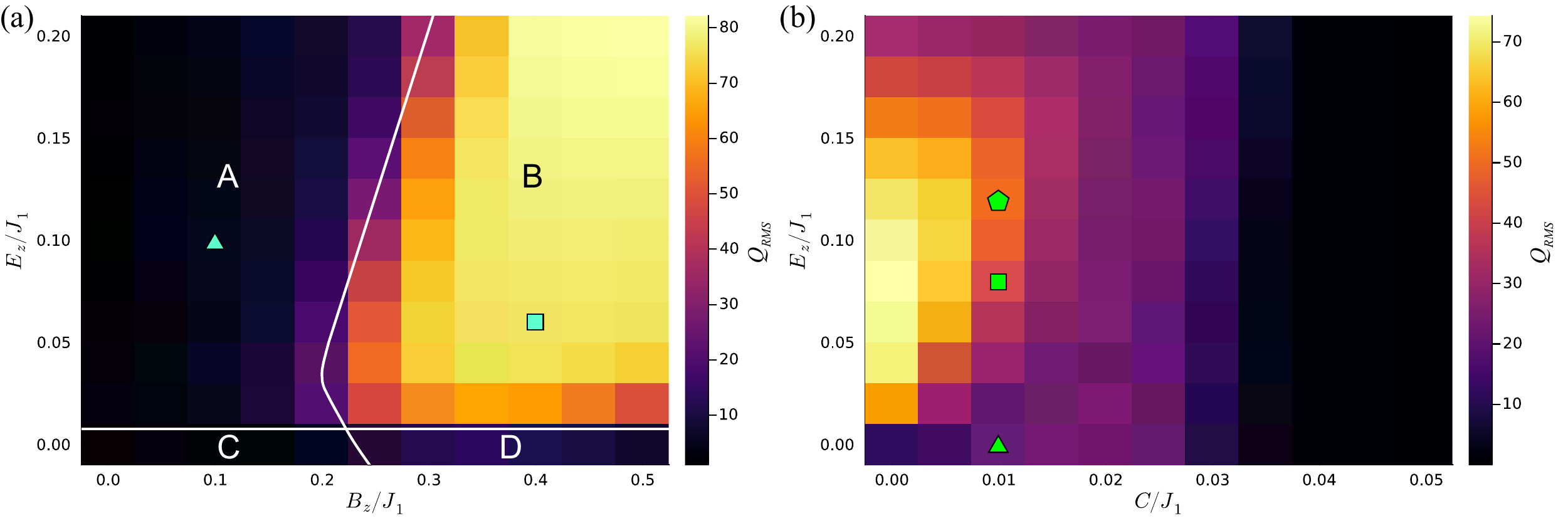}
	\caption{\label{efieldphase} Effect of an external electric field on the spin textures. (a) Phase diagram  illustrating the evolution of $Q_{RMS}$ as a function of perpendicular magnetic field and electric fields. Phase A is an out of plane spin spiral phase, phase B is a N\'eel skyrmion lattice, phase C is an in-plane spin spiral phase and phase D an array of mixed skyrmions. The turquoise triangle and square indicate the $(B_z,E_z)$ coordinates of the spin textures shown in Fig.~\ref{efieldimgs}(a) and (b), respectively. (b) Interplay between electric field and dipole-dipole interactions in determining the eventual spin texture. $B_z/J_1$ is fixed at 0.3. The green triangle, square and pentagon correspond to the series of snapshots shown in Fig.~\ref{ezdipoleimgs}(a-c), respectively, showing a transition between Bloch and N\'eel skyrmion types as $E_z$ increases. $J_2 = -0.3$, $J_3 = -0.25$, $K = 0$ and $T = 0$ for both phase diagrams.
	}
\end{figure*}

For our calculations we reduce the units in terms of $E_z = EP_0 / J_1$. In Fig.~\ref{efieldphase}(a) we present the phase diagram dependent on magnetic and electric fields, with no dipole-dipole interaction. Phase C is the same spin spiral phase from Fig.~\ref{j2j3phase}, and Phase D is the same skyrmion phase. Phase A is also a spin spiral phase, however the spins rotate out of plane instead of in-plane. Phase B is a high density N\'eel skyrmion crystal. Snapshots of the new phases are shown in Fig.~\ref{efieldimgs}. Even in very small electric fields, we observe the near complete removal of antiskyrmions, leaving a pure N\'eel skyrmion phase with almost no transition region. With increasing electric field, the skyrmion concentration rapidly increases and forms a N\'eel skyrmion lattice. 
	
If dipole-dipole interactions are included, there is a competition between N\'eel and Bloch skrymions because the electric field favors N\'eel skyrmions, while the dipole-dipole interaction favors Bloch skyrmions. This interplay between electric field and dipole-dipole interactions is presented in Fig.~\ref{efieldphase}(b). If the dipole-dipole interaction is small, we find only Bloch skyrmions, shown in Fig.~\ref{ezdipoleimgs}(a). As the electric field is increased, while keeping the dipole-dipole interaction constant, a transition to intermediate helicity N\'eel/Bloch skyrmions occurs first [Fig.~\ref{ezdipoleimgs}(b)], with uniform N\'eel skyrmions eventually emerging at higher $E_z$ as shown in Fig.~\ref{ezdipoleimgs}(c). The sequence of spin textures highlighted in these images is indicated by the green polygons in Fig.~\ref{efieldphase}(b). Crucially, an intermediate helicity $\varphi_0\approx -\pi/4$ is observed across the entire skyrmion array in Fig.~\ref{ezdipoleimgs}(b). This suggests that an ``electric annealing'' process featuring a gradual reduction of $E_z$ to zero may permit the stabilization of a Bloch skyrmion array with a single homogeneous helicity, despite the $\pm\pi/2$ degeneracy - thus delivering full-spectrum $2\pi$ helicity selectivity. In other words, the helicity should exhibit a hysteresis effect as a function of external electric field [Fig.~\ref{ezdipoleimgs}(a,b)].  Moreover, the progressive nature of our $E_z$-induced helicity switching differs qualitatively from previous simulations~\cite{Yao2020}, promising fine control of the $\varphi_0$-dependent potential well for qubit applications. 

\begin{figure}[b!]
    \caption{\label{efieldimgs} Tunable spin textures observed upon varying perpendicular electric and magnetic fields, as shown in Fig.~\ref{efieldphase}(a). (a) Phase A (turquoise triangle) is an out of plane spin spiral phase stabilized at finite dipole-dipole coupling and low magnetic fields. (b) Phase C (turquoise square) is a N\'eel skyrmion lattice. $J_2 = -0.3$, $J_3 = -0.25$, $K = 0$ and $T = 0$ for both textures.
	}
\end{figure}
\begin{figure}[b!]
	\includegraphics[clip=true, width=0.95\columnwidth]{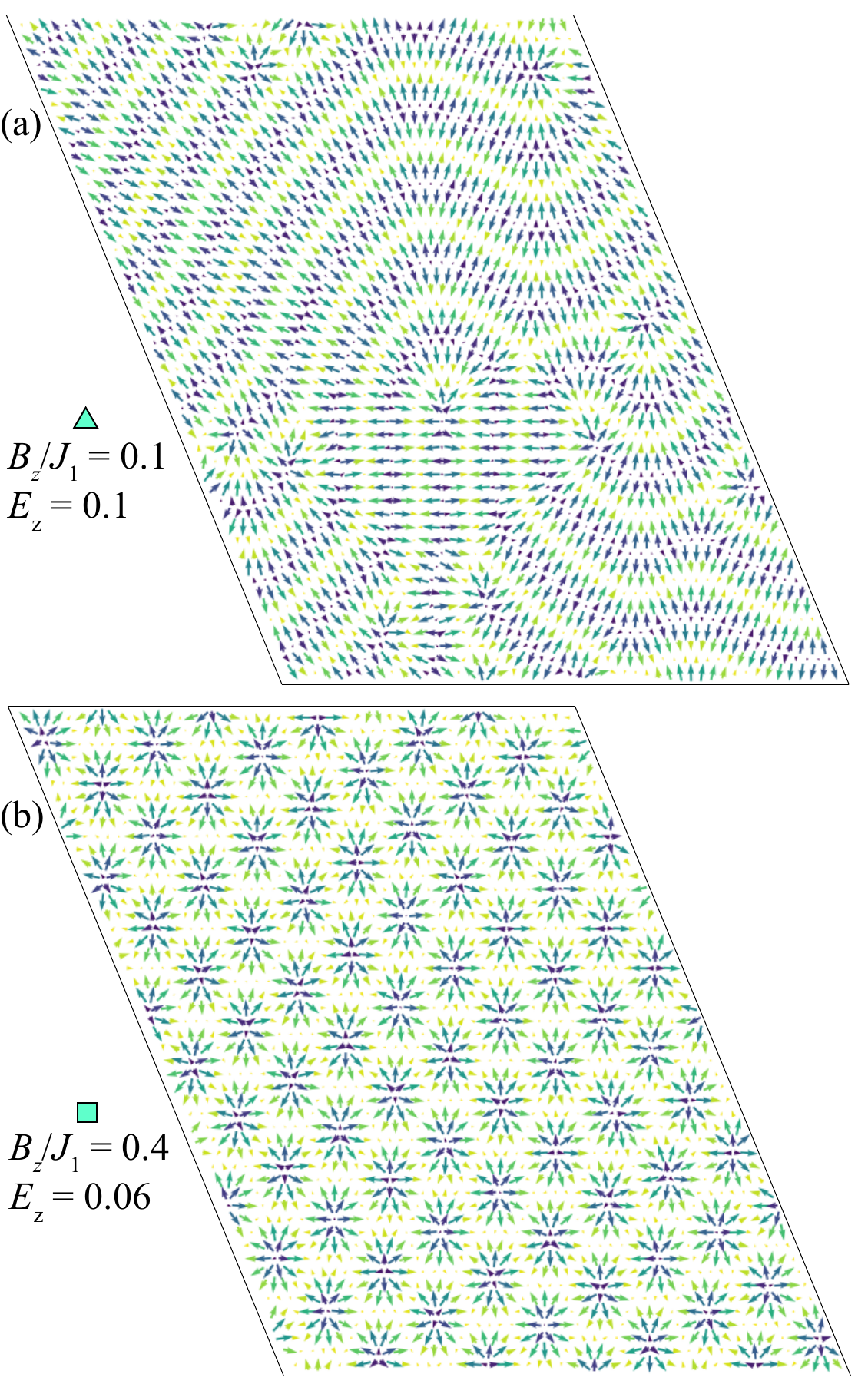}
\end{figure}

\begin{figure}[t]
	\includegraphics[clip=true, width=0.95\columnwidth]{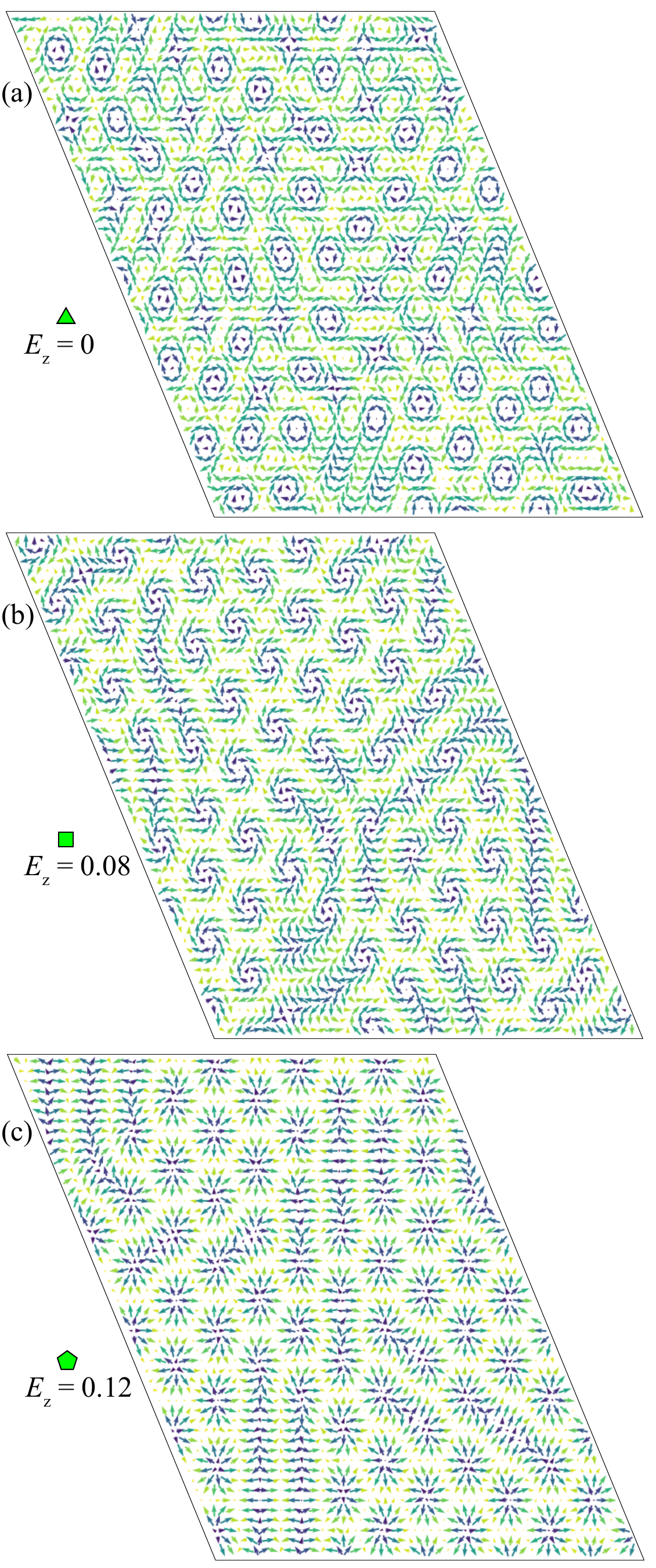}
\end{figure}
\begin{figure}[t]
    \caption{\label{ezdipoleimgs} Electrically-induced transition between Bloch and N\'eel skyrmions as illustrated in Fig.~\ref{efieldphase}(b). (a). At zero electric field [green triangle in Fig.~\ref{efieldphase}(b)], and a small dipole-dipole interaction, a dense skyrmion phase with predominant Bloch helicity is formed. (b) In small electric fields $E_z=0.08$ [green square in Fig.~\ref{efieldphase}(b)], intermediate helicity N\'eel/Bloch skyrmions with $-\pi/2<\varphi_0<0$ are formed. The small population of antiskyrmions which was present for $E_z=0$ is quenched.  (c) Upon further increasing the electric field to $E_z=0.12$, $\varphi_0\rightarrow0$ creating a mixture of skyrmions and stripes with uniform N\'eel helicity. Here, a skyrmion crystal [e.g. Fig.~\ref{efieldimgs}(b)] can be easily recovered by increasing $B_z$. All three textures are calculated using $J_2/J_1 = -0.3$, $J_3/J_1 = -0.25$, $K=0$, $B_z/J_1 = 0.3$, $T = 0$ and $C/J_1 = 0.01$.
	}
\end{figure}

\begin{figure}[b!]
	\includegraphics[clip=true, width=0.95\columnwidth]{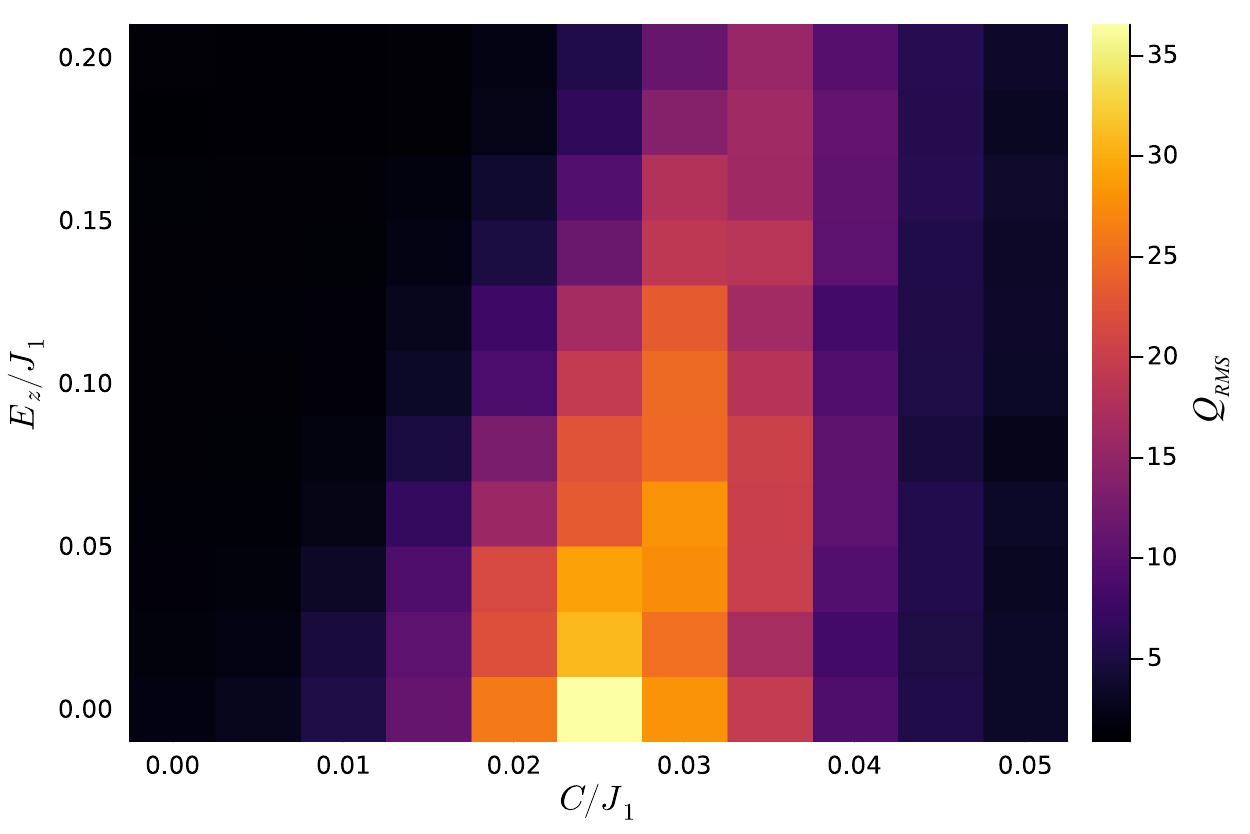}
    \caption{\label{ZFEz} Evolution of topological charge density with perpendicular electric fields and dipole-dipole interaction. The phase diagram is calculated using $J_2/J_1 = -0.3$, $J_3/J_1 = -0.25$, $K=0$, $B_z/J_1 = 0$ and $T = 0$.
	}
\end{figure}

Finally, we examine the electric field evolution of the technologically-relevant Bloch skyrmion crystal stabilized in zero magnetic field [as depicted in Fig.~\ref{bzdipoleimages}(a)]. Figure~\ref{ZFEz} illustrates the variation in $Q_{RMS}$ upon varying $E_z$ and the dipole-dipole interaction strength $C$. Here, the Bloch crystal corresponds to the peak in $Q_{RMS}$ at $C/J_1=0.025$ and $E_z=0$. As $E_z$ increases, the topological charge density falls rapidly: this implies that in contrast to the Bloch crystal, the N\'eel skyrmion lattice is only stable in finite perpendicular magnetic fields.  

\section{Conclusions}
For quantum computing based on skyrmions, as well as skyrmion memory devices, control of skyrmion helicity and spatial configuration is paramount. We have computationally studied centrosymmetric triangular 2D lattice systems, finding rich spin texture phases due to magnetic frustration. We model the RKKY interaction using exchange interactions, by considering first, second, and third nearest neighbor Heisenberg exchange terms. We select values for the second and third nearest neighbor interaction parameters by qualitatively matching the emergent skyrmion phase to the experimental phase diagrams for Gd$_2$PdSi$_3$ from Refs.~\cite{Gomil25, Takashi19}, as shown in Fig.~\ref{tempbzphase}. We find that in systems with no other skyrmion interactions or applied fields, phases with mixed skyrmion types and antiskyrmions exist, as shown in Fig.~\ref{j2j3images}. We next added easy axis anisotropy, and have confirmed that it helps to stabilize and expand skyrmion phases, including increasing the skyrmion density (see Fig.~\ref{bzani}).
	
We then explicitly included a magnetic dipole-dipole term in the Hamiltonian (Eq.~\ref{dipole}). By introducing magnetic dipole-dipole interactions, we have demonstrated that the interaction favors Bloch skyrmions over N\'eel skymions and antiskyrmions. Crucially, we have found that the dipole-dipole interaction can stabilize a low or zero magnetic field skyrmion lattice phase. This is important for practical applications since the requirement for a high field superconducting magnet is eliminated, potentially making skyrmion-based devices considerably less expensive. Furthermore, at high dipole-dipole interaction strengths, a Bloch meron/antimeron lattice phase can be formed in the absence of any DMI (see Figs.~\ref{bzdipolephase} and~\ref{bzdipoleimages}). 
		
To control the skyrmion type, and recover N\'eel skyrmions, we have studied electric field-induced phase transitions due to the inhomogeneous magnetoelectric effect. We have found that if dipole-dipole interactions are negligible, applied electric field nearly instantaneously forces all skyrmions to adopt N\'eel helicity, thus stabilizing a N\'eel skyrmion lattice [see Fig.~\ref{efieldphase}(a)]. The skyrmion density increases rapidly even with moderate applied electric fields. However, if dipole-dipole interactions are included, there is a competition where dipole-dipole interactions favor Bloch skyrmions, and electric field favors N\'eel skyrmions. This interplay is shown in Figs.~\ref{efieldphase}(b) and~\ref{ezdipoleimgs}: a transition from Bloch to N\'eel skyrmions can be forced by applying perpendicular electric fields, with a mixed helicity N\'eel/Bloch intermediate phase formed. For qubits based on a skyrmion-induced topological phase transition in a proximate superconductor, N\'eel helicity is desirable for strong coupling to a superconducting vortex; therefore, using a locally applied electric field to create N\'eel skyrmions may prove invaluable for nucleating and braiding Majorana zero modes. For qubits based on skyrmion two-level systems with helicity quantization, Bloch skyrmions are preferred; therefore, materials with moderately strong dipole-dipole interactions compared to the exchange ($C/J_1\sim0.025$) would be desirable. In practice, this means identifying materials with large local moments (e.g. 4$f$ rare earth systems) and a relatively large lattice parameter or spatial separation between the magnetic moments. The results presented in this work hence provide a roadmap for future \textit{ab initio} design of noncollinear magnetic materials for quantum applications.

\section*{Acknowledgements}
	
R.R. and Y.D. are supported by National Science Foundation (NSF) grant No. 2228841.  A.P.P. is supported by NSF LEAPS-MPS Award No. 2419041.
\\
\section*{Data Availability}
The data that support the findings of this study are available within the main text of this article. Any other relevant data are available from the corresponding authors upon reasonable request.
	
\bibliography{refs}

@article{Bogdanov89,
	author = {Bogdanov, A. and Yablonskii, D.},
	year = {1989},
	month = {01},
	pages = {101},
	title = {Thermodynamically stable "vortices" in magnetically ordered crystals. The mixed state of magnets},
	volume = {68},
	journal = {Sov. Phys. JETP}
}

@article{
	Muhlbauer09,
	author = {S. Mühlbauer  and B. Binz  and F. Jonietz  and C. Pfleiderer  and A. Rosch  and A. Neubauer  and R. Georgii  and P. Böni },
	title = {Skyrmion Lattice in a Chiral Magnet},
	doi = {10.1126/science.1166767},
    journal = {Science},
	volume = {323},
	number = {5916},
	pages = {915-919},
	year = {2009}
	}

@article{
	Seki12,
	author = {S. Seki  and X. Z. Yu  and S. Ishiwata  and Y. Tokura },
	title = {Observation of Skyrmions in a Multiferroic Material},
	journal = {Science},
	volume = {336},
	number = {6078},
	pages = {198-201},
	year = {2012},
	doi = {10.1126/science.1214143},
}

@article{Back20,
	doi = {10.1088/1361-6463/ab8418},
	url = {https://doi.org/10.1088/1361-6463/ab8418},
	year = {2020},
	month = {jun},
	publisher = {IOP Publishing},
	volume = {53},
	number = {36},
	pages = {363001},
	author = {Back, C and Cros, V and Ebert, H and Everschor-Sitte, K and Fert, A and Garst, M and Ma, Tianping and Mankovsky, S and Monchesky, T L and Mostovoy, M and Nagaosa, N and Parkin, S S P and Pfleiderer, C and Reyren, N and Rosch, A and Taguchi, Y and Tokura, Y and von Bergmann, K and Zang, Jiadong},
	title = {The 2020 skyrmionics roadmap},
	journal = {Journal of Physics D: Applied Physics}
}

@article{Li23,
	author = {Li, Sheng and Wang, Xuewen and Rasing, Theo},
	title = {Magnetic skyrmions: Basic properties and potential applications},
	journal = {Interdisciplinary Materials},
	volume = {2},
	number = {2},
	pages = {260-289},
	doi = {https://doi.org/10.1002/idm2.12072},
	year = {2023}
}

@article{Khatua23,
	author = {Khatua, J. and Sana, Biprojit and Zorko, A. and Gomilšek, M. and Sethupathi, Kanikrishnan and Rao, M S and Baenitz, M. and Schmidt, B. and Khuntia, Panchanan},
	year = {2023},
	month = {11},
	pages = {1-60},
	title = {Experimental signatures of quantum and topological states in frustrated magnetism},
	volume = {1041},
	journal = {Physics Reports},
	doi = {10.1016/j.physrep.2023.09.008}
}

@article{Dong23,
	title = {Strain-tuning Bloch- and Néel-type magnetic skyrmions: A phase-field simulation},
	journal = {Scripta Materialia},
	volume = {222},
	pages = {114994},
	year = {2023},
	issn = {1359-6462},
	doi = {https://doi.org/10.1016/j.scriptamat.2022.114994},
	url = {https://www.sciencedirect.com/science/article/pii/S1359646222004894},
	author = {Shouzhe Dong and Jing Wang and Xiaoming Shi and Deshan Liang and Hasnain Mehdi Jafri and Chengchao Hu and Ke Jin and Houbing Huang}
}

@Article{Litzius20,
	author={Litzius, Kai
	and Leliaert, Jonathan
	and Bassirian, Pedram
	and Rodrigues, Davi
	and Kromin, Sascha
	and Lemesh, Ivan
	and Zazvorka, Jakub
	and Lee, Kyu-Joon
	and Mulkers, Jeroen
	and Kerber, Nico
	and Heinze, Daniel
	and Keil, Niklas
	and Reeve, Robert M.
	and Weigand, Markus
	and Van Waeyenberge, Bartel
	and Sch{\"u}tz, Gisela
	and Everschor-Sitte, Karin
	and Beach, Geoffrey S. D.
	and Kl{\"a}ui, Mathias},
	title={The role of temperature and drive current in skyrmion dynamics},
	journal={Nature Electronics},
	year={2020},
	month={Jan},
	day={01},
	volume={3},
	number={1},
	pages={30-36},
	issn={2520-1131},
	doi={10.1038/s41928-019-0359-2},
	url={https://doi.org/10.1038/s41928-019-0359-2}
}

@article{Shustin23,
	title = {Higher-order magnetic skyrmions in nonuniform magnetic fields},
	author = {Shustin, M. S. and Stepanenko, V. A. and Dzebisashvili, D. M.},
	journal = {Phys. Rev. B},
	volume = {107},
	issue = {19},
	pages = {195428},
	numpages = {18},
	year = {2023},
	month = {May},
	publisher = {American Physical Society},
	doi = {10.1103/PhysRevB.107.195428},
	url = {https://link.aps.org/doi/10.1103/PhysRevB.107.195428}
}

@article{Apostoloff24,
	title = {Deformation of a N{\'e}el-type skyrmion in a weak inhomogeneous magnetic field: Magnetization Ansatz and interaction with a Pearl vortex},
	author = {Apostoloff, S. S. and Andriyakhina, E. S. and Burmistrov, I. S.},
	journal = {Phys. Rev. B},
	volume = {109},
	issue = {10},
	pages = {104406},
	numpages = {16},
	year = {2024},
	month = {Mar},
	publisher = {American Physical Society},
	doi = {10.1103/PhysRevB.109.104406},
	url = {https://link.aps.org/doi/10.1103/PhysRevB.109.104406}
}

@article{Pyatakov15,
	title = {Spin flexoelectricity and chiral spin structures in magnetic films},
	journal = {Journal of Magnetism and Magnetic Materials},
	volume = {383},
	pages = {255-258},
	year = {2015},
	note = {Selected papers from the sixth Moscow International Symposium on Magnetism (MISM-2014)},
	issn = {0304-8853},
	doi = {https://doi.org/10.1016/j.jmmm.2014.11.035},
	url = {https://www.sciencedirect.com/science/article/pii/S0304885314011433},
	author = {A.P. Pyatakov and A.S. Sergeev and F.A. Mikailzade and A.K. Zvezdin}
}

@article{Wang19,
	title = {Thermally assisted skyrmion drag in a nonuniform electric field},
	author = {Wang, Xi-Guang and Chotorlishvili, L. and Guo, Guang-Hua and Jia, C.-L. and Berakdar, J.},
	journal = {Phys. Rev. B},
	volume = {99},
	issue = {6},
	pages = {064426},
	numpages = {10},
	year = {2019},
	month = {Feb},
	publisher = {American Physical Society},
	doi = {10.1103/PhysRevB.99.064426},
	url = {https://link.aps.org/doi/10.1103/PhysRevB.99.064426}
}

@article{Wang22,
	author = {Wang, Ya-Dong and Wei, Zhi-Jian and Tu, Hao-Ran and Zhang, Chen-Hui and Hou, Zhi-Peng},
	title = {Electric field manipulation of magnetic skyrmions},
	journal = {Rare Metals},
	volume = {41},
	number = {12},
	pages = {4000-4014},
	doi = {https://doi.org/10.1007/s12598-022-02084-0},
	year = {2022}
}

@article{Lancaster19,
	author = {Tom Lancaster},
	title = {Skyrmions in magnetic materials},
	journal = {Contemporary Physics},
	volume = {60},
	number = {3},
	pages = {246--261},
	year = {2019},
	publisher = {Taylor \& Francis},
	doi = {10.1080/00107514.2019.1699352}	
	}

@article{Morikawa13,
	title = {Crystal chirality and skyrmion helicity in MnSi and (Fe, Co)Si as determined by transmission electron microscopy},
	author = {Morikawa, D. and Shibata, K. and Kanazawa, N. and Yu, X. Z. and Tokura, Y.},
	journal = {Phys. Rev. B},
	volume = {88},
	issue = {2},
	pages = {024408},
	numpages = {4},
	year = {2013},
	month = {Jul},
	publisher = {American Physical Society},
	doi = {10.1103/PhysRevB.88.024408},
	url = {https://link.aps.org/doi/10.1103/PhysRevB.88.024408}
}

@article{Gomil25,
	title = {Anisotropic Skyrmion and Multi-$q$ Spin Dynamics in Centrosymmetric {{Gd$_2$PdSi$_3$}}},
	author = {Gomil\ifmmode \check{s}\else \v{s}\fi{}ek, M. and Hicken, T. J. and Wilson, M. N. and Franke, K. J. A. and Huddart, B. M. and \ifmmode \check{S}\else \v{S}\fi{}tefan\ifmmode \check{c}\else \v{c}\fi{}i\ifmmode \check{c}\else \v{c}\fi{}, A. and Holt, S. J. R. and Balakrishnan, G. and Mayoh, D. A. and Birch, M. T. and Moody, S. H. and Luetkens, H. and Guguchia, Z. and Telling, M. T. F. and Baker, P. J. and Clark, S. J. and Lancaster, T.},
	journal = {Phys. Rev. Lett.},
	volume = {134},
	issue = {4},
	pages = {046702},
	numpages = {8},
	year = {2025},
	month = {Jan},
	publisher = {American Physical Society},
	doi = {10.1103/PhysRevLett.134.046702},
	url = {https://link.aps.org/doi/10.1103/PhysRevLett.134.046702}
}

@article{Takashi19,
	author = {Takashi Kurumaji  and Taro Nakajima  and Max Hirschberger  and Akiko Kikkawa  and Yuichi Yamasaki  and Hajime Sagayama  and Hironori Nakao  and Yasujiro Taguchi  and {Taka-hisa} Arima  and Yoshinori Tokura },
	title = {Skyrmion lattice with a giant topological Hall effect in a frustrated triangular-lattice magnet},
	journal = {Science},
	volume = {365},
	number = {6456},
	pages = {914-918},
	year = {2019},
	doi = {10.1126/science.aau0968}
}

@article{Hirschberger20,
	title = {High-field depinned phase and planar Hall effect in the skyrmion host {{Gd$_2$PdSi$_3$}}},
	author = {Hirschberger, Max and Nakajima, Taro and Kriener, Markus and Kurumaji, Takashi and Spitz, Leonie and Gao, Shang and Kikkawa, Akiko and Yamasaki, Yuichi and Sagayama, Hajime and Nakao, Hironori and Ohira-Kawamura, Seiko and Taguchi, Yasujiro and Arima, Taka-hisa and Tokura, Yoshinori},
	journal = {Phys. Rev. B},
	volume = {101},
	issue = {22},
	pages = {220401(R)},
	numpages = {6},
	year = {2020},
	month = {Jun},
	publisher = {American Physical Society},
	doi = {10.1103/PhysRevB.101.220401},
	url = {https://link.aps.org/doi/10.1103/PhysRevB.101.220401}
}

@article{Zhang20,
	doi = {10.1088/1367-2630/aba650},
	url = {https://doi.org/10.1088/1367-2630/aba650},
	year = {2020},
	month = {aug},
	publisher = {IOP Publishing},
	volume = {22},
	number = {8},
	pages = {083056},
	author = {Zhang, Han and Huang, Qing and Hao, Lin and Yang, Junyi and Noordhoek, Kyle and Pandey, Shashi and Zhou, Haidong and Liu, Jian},
	title = {Anomalous magnetoresistance in centrosymmetric skyrmion-lattice magnet {{Gd$_2$PdSi$_3$}}},
	journal = {New Journal of Physics}
}

@article{Takagi22,
	author={Takagi, Rina and Matsuyama, Naofumi	and Ukleev, Victor and Yu, Le and White, Jonathan S. and Francoual, Sonia and Mardegan, Jos{\'e} R. L. and Hayami, Satoru	and Saito, Hiraku and Kaneko, Koji and Ohishi, Kazuki and {\={O}}nuki, Yoshichika and Arima, Taka-hisa and Tokura, Yoshinori and Nakajima, Taro and Seki, Shinichiro},
	title={Square and rhombic lattices of magnetic skyrmions in a centrosymmetric binary compound},
	doi = {10.1038/s41467-022-29131-9},
    journal={Nature Communications},
	year={2022},
	month={Mar},
	day={30},
	volume={13},
	number={1},
	pages={1472},
}

@article{Grant26,
	title = {Surface morphology and electronic properties of cleaved centrosymmetric {{EuAl$_4$}}},
	author = {Grant, Jarred L. and King, David and Vlastos, Nikiphoros and Rivlis, Raz and Carter, Jefferson A. and Dahnovsky, Yuri and Tang, Jinke and Chien, TeYu},
	journal = {Phys. Rev. B},
	volume = {113},
	issue = {7},
	pages = {075418},
	numpages = {8},
	year = {2026},
	month = {Feb},
	publisher = {American Physical Society},
	doi = {10.1103/6z6c-glsq},
	url = {https://link.aps.org/doi/10.1103/6z6c-glsq}
}

@article{Shang21,
	title = {Anomalous Hall resistivity and possible topological Hall effect in the {{EuAl$_4$}} antiferromagnet},
	author = {Shang, T. and Xu, Y. and Gawryluk, D. J. and Ma, J. Z. and Shiroka, T. and Shi, M. and Pomjakushina, E.},
	journal = {Phys. Rev. B},
	volume = {103},
	issue = {2},
	pages = {L020405},
	numpages = {7},
	year = {2021},
	month = {Jan},
	publisher = {American Physical Society},
	doi = {10.1103/PhysRevB.103.L020405},
	url = {https://link.aps.org/doi/10.1103/PhysRevB.103.L020405}
}

@article{Okubo12,
	title = {Multiple-$q$ States and the Skyrmion Lattice of the Triangular-Lattice Heisenberg Antiferromagnet under Magnetic Fields},
	author = {Okubo, Tsuyoshi and Chung, Sungki and Kawamura, Hikaru},
	journal = {Phys. Rev. Lett.},
	volume = {108},
	issue = {1},
	pages = {017206},
	numpages = {5},
	year = {2012},
	month = {Jan},
	publisher = {American Physical Society},
	doi = {10.1103/PhysRevLett.108.017206},
	url = {https://link.aps.org/doi/10.1103/PhysRevLett.108.017206}
}

@Article{Leonov15,
	author={Leonov, A. O.
	and Mostovoy, M.},
	title={Multiply periodic states and isolated skyrmions in an anisotropic frustrated magnet},
	journal={Nature Communications},
	year={2015},
	month={Sep},
	day={23},
	volume={6},
	number={1},
	pages={8275},
	issn={2041-1723},
	doi={10.1038/ncomms9275},
	url={https://doi.org/10.1038/ncomms9275}
}

@article{Lohani19,
	title = {Quantum Skyrmions in Frustrated Ferromagnets},
	author = {Lohani, Vivek and Hickey, Ciar\'an and Masell, Jan and Rosch, Achim},
	journal = {Phys. Rev. X},
	volume = {9},
	issue = {4},
	pages = {041063},
	numpages = {14},
	year = {2019},
	month = {Dec},
	publisher = {American Physical Society},
	doi = {10.1103/PhysRevX.9.041063},
	url = {https://link.aps.org/doi/10.1103/PhysRevX.9.041063}
}

@article{Paddison22,
	title = {Magnetic Interactions of the Centrosymmetric Skyrmion Material {{Gd$_2$PdSi$_3$}}},
	author = {Paddison, Joseph A. M. and Rai, Binod K. and May, Andrew F. and Calder, Stuart and Stone, Matthew B. and Frontzek, Matthias D. and Christianson, Andrew D.},
	journal = {Phys. Rev. Lett.},
	volume = {129},
	issue = {13},
	pages = {137202},
	numpages = {6},
	year = {2022},
	month = {Sep},
	publisher = {American Physical Society},
	doi = {10.1103/PhysRevLett.129.137202},
	url = {https://link.aps.org/doi/10.1103/PhysRevLett.129.137202}
}

@article{Dong24,
	title = {Fermi Surface Nesting Driving the RKKY Interaction in the Centrosymmetric Skyrmion Magnet {{Gd$_2$PdSi$_3$}}},
	author = {Dong, Yuyang and Arai, Yosuke and Kuroda, Kenta and Ochi, Masayuki and Tanaka, Natsumi and Wan, Yuxuan and Watson, Matthew D. and Kim, Timur K. and Cacho, Cephise and Hashimoto, Makoto and Lu, Donghui and Aoki, Yuji and Matsuda, Tatsuma D. and Kondo, Takeshi},
	journal = {Phys. Rev. Lett.},
	volume = {133},
	issue = {1},
	pages = {016401},
	numpages = {6},
	year = {2024},
	month = {Jul},
	publisher = {American Physical Society},
	doi = {10.1103/PhysRevLett.133.016401},
	url = {https://link.aps.org/doi/10.1103/PhysRevLett.133.016401}
}

@article{Matsuyama23,
	title = {Quantum oscillations in the centrosymmetric skyrmion-hosting magnet {{GdRu$_2$Si$_2$}}},
	author = {Matsuyama, N. and Nomura, T. and Imajo, S. and Nomoto, T. and Arita, R. and Sudo, K. and Kimata, M. and Khanh, N. D. and Takagi, R. and Tokura, Y. and Seki, S. and Kindo, K. and Kohama, Y.},
	journal = {Phys. Rev. B},
	volume = {107},
	issue = {10},
	pages = {104421},
	numpages = {10},
	year = {2023},
	month = {Mar},
	publisher = {American Physical Society},
	doi = {10.1103/PhysRevB.107.104421},
	url = {https://link.aps.org/doi/10.1103/PhysRevB.107.104421}
}

@Article{Heinze11,
	author={Heinze, Stefan
	and von Bergmann, Kirsten
	and Menzel, Matthias
	and Brede, Jens
	and Kubetzka, Andr{\'e}
	and Wiesendanger, Roland
	and Bihlmayer, Gustav
	and Bl{\"u}gel, Stefan},
	title={Spontaneous atomic-scale magnetic skyrmion lattice in two dimensions},
	journal={Nature Physics},
	year={2011},
	month={Sep},
	day={01},
	volume={7},
	number={9},
	pages={713-718},
	issn={1745-2481},
	doi={10.1038/nphys2045},
	url={https://doi.org/10.1038/nphys2045}
}

@article{Katsura05,
	title = {Spin Current and Magnetoelectric Effect in Noncollinear Magnets},
	author = {Katsura, Hosho and Nagaosa, Naoto and Balatsky, Alexander V.},
	journal = {Phys. Rev. Lett.},
	volume = {95},
	issue = {5},
	pages = {057205},
	numpages = {4},
	year = {2005},
	month = {Jul},
	publisher = {American Physical Society},
	doi = {10.1103/PhysRevLett.95.057205},
	url = {https://link.aps.org/doi/10.1103/PhysRevLett.95.057205}
}

@article{Nikolaev19,
	title = {Microscopic theory of electric polarization induced by skyrmionic order in {{GaV$_4$S$_8$}}},
	author = {Nikolaev, S. A. and Solovyev, I. V.},
	journal = {Phys. Rev. B},
	volume = {99},
	issue = {10},
	pages = {100401(R)},
	numpages = {5},
	year = {2019},
	month = {Mar},
	publisher = {American Physical Society},
	doi = {10.1103/PhysRevB.99.100401},
	url = {https://link.aps.org/doi/10.1103/PhysRevB.99.100401}
}

@article{Jefremovas25,
	title = {The role of magnetic dipolar interactions in skyrmion lattices},
	journal = {Newton},
	volume = {1},
	number = {2},
	pages = {100036},
	year = {2025},
	issn = {2950-6360},
	doi = {https://doi.org/10.1016/j.newton.2025.100036},
	url = {https://www.sciencedirect.com/science/article/pii/S2950636025000283},
	author = {Elizabeth M. Jefremovas and Kilian Leutner and Miriam G. Fischer and Jorge Marqués-Marchán and Thomas B. Winkler and Agustina Asenjo and Jairo Sinova and Robert Frömter and Mathias Kläui}
}

@article{Hastings70,
	author = {Hastings, W. K.},
	title = {Monte Carlo sampling methods using Markov chains and their applications},
	journal = {Biometrika},
	volume = {57},
	number = {1},
	pages = {97-109},
	year = {1970},
	month = {04},
	issn = {0006-3444},
	doi = {10.1093/biomet/57.1.97}
}

@article{Psaroudaki21,
  title = {Skyrmion Qubits: A New Class of Quantum Logic Elements Based on Nanoscale Magnetization},
  author = {Psaroudaki, Christina and Panagopoulos, Christos},
  journal = {Phys. Rev. Lett.},
  volume = {127},
  issue = {6},
  pages = {067201},
  numpages = {6},
  year = {2021},
  month = {Aug},
  publisher = {American Physical Society},
  doi = {10.1103/PhysRevLett.127.067201},
  url = {https://link.aps.org/doi/10.1103/PhysRevLett.127.067201}
}

@Article{Bogdanov20,
author={Bogdanov, Alexei N.
and Panagopoulos, Christos},
title={Physical foundations and basic properties of magnetic skyrmions},
journal={Nature Reviews Physics},
year={2020},
month={Sep},
day={01},
volume={2},
number={9},
pages={492-498},
issn={2522-5820},
doi={10.1038/s42254-020-0203-7},
url={https://doi.org/10.1038/s42254-020-0203-7}
}

@article{Luo21,
    author = {Luo, Shijiang and You, Long},
    title = {Skyrmion devices for memory and logic applications},
    journal = {APL Materials},
    volume = {9},
    number = {5},
    pages = {050901},
    year = {2021},
    month = {05},
    issn = {2166-532X},
    doi = {10.1063/5.0042917}
}

@article{Rex19,
  title = {Majorana bound states in magnetic skyrmions imposed onto a superconductor},
  author = {Rex, Stefan and Gornyi, Igor V. and Mirlin, Alexander D.},
  journal = {Phys. Rev. B},
  volume = {100},
  issue = {6},
  pages = {064504},
  numpages = {13},
  year = {2019},
  month = {Aug},
  publisher = {American Physical Society},
  doi = {10.1103/PhysRevB.100.064504},
  url = {https://link.aps.org/doi/10.1103/PhysRevB.100.064504}
}

@article{Nothhelfer22,
  title = {Steering Majorana braiding via skyrmion-vortex pairs: A scalable platform},
  author = {Nothhelfer, Jonas and D\'{\i}az, Sebasti\'an A. and Kessler, Stephan and Meng, Tobias and Rizzi, Matteo and Hals, Kjetil M. D. and Everschor-Sitte, Karin},
  journal = {Phys. Rev. B},
  volume = {105},
  issue = {22},
  pages = {224509},
  numpages = {6},
  year = {2022},
  month = {Jun},
  publisher = {American Physical Society},
  doi = {10.1103/PhysRevB.105.224509},
  url = {https://link.aps.org/doi/10.1103/PhysRevB.105.224509}
}

@article{Konakanchi23,
  title = {Platform for braiding Majorana modes with magnetic skyrmions},
  author = {Konakanchi, Shiva T. and V\"ayrynen, Jukka I. and Chen, Yong P. and Upadhyaya, Pramey and Rokhinson, Leonid P.},
  journal = {Phys. Rev. Res.},
  volume = {5},
  issue = {3},
  pages = {033109},
  numpages = {12},
  year = {2023},
  month = {Aug},
  publisher = {American Physical Society},
  doi = {10.1103/PhysRevResearch.5.033109},
  url = {https://link.aps.org/doi/10.1103/PhysRevResearch.5.033109}
}

@article{Nagaosa13,
author={Nagaosa, Naoto
and Tokura, Yoshinori},
title={Topological properties and dynamics of magnetic skyrmions},
journal={Nature Nanotechnology},
year={2013},
month={Dec},
day={01},
volume={8},
number={12},
pages={899-911},
doi={10.1038/nnano.2013.243},
}

@article{BLY83,
  title={Inhomogeneous magnetoelectric effect},
  author={Viktor Grigorievich Baryakhtar and Victor A. L'vov and D. A. Yablonskii},
  journal={Jetp Letters},
  year={1983},
  volume={37},
  pages={673},
  url={https://api.semanticscholar.org/CorpusID:115936765}
}

@Article{Fujishiro21,
author={Fujishiro, Yukako
and Kanazawa, Naoya
and Kurihara, Ryosuke
and Ishizuka, Hiroaki
and Hori, Tomohiro
and Yasin, Fehmi Sami
and Yu, Xiuzhen
and Tsukazaki, Atsushi
and Ichikawa, Masakazu
and Kawasaki, Masashi
and Nagaosa, Naoto
and Tokunaga, Masashi
and Tokura, Yoshinori},
title={Giant anomalous Hall effect from spin-chirality scattering in a chiral magnet},
journal={Nature Communications},
year={2021},
month={Jan},
day={12},
volume={12},
number={1},
pages={317},
issn={2041-1723},
doi={10.1038/s41467-020-20384-w},
url={https://doi.org/10.1038/s41467-020-20384-w}
}

@book{Jackson99,
  address = {New York, {NY}},
  author = {Jackson, John David},
  edition = {3rd ed.},
  isbn = {9780471309321},
  lccn = {538.3537.8},
  publisher = {Wiley},
  title = {Classical electrodynamics},
  url = {http://cdsweb.cern.ch/record/490457},
  year = 1999
}

@Article{Yu18,
author={Yu, X. Z.
and Koshibae, W.
and Tokunaga, Y.
and Shibata, K.
and Taguchi, Y.
and Nagaosa, N.
and Tokura, Y.},
title={Transformation between meron and skyrmion topological spin textures in a chiral magnet},
journal={Nature},
year={2018},
month={Dec},
day={01},
volume={564},
number={7734},
pages={95-98},
issn={1476-4687},
doi={10.1038/s41586-018-0745-3},
url={https://doi.org/10.1038/s41586-018-0745-3}
}

@article{Hayami21,
  title = {Meron-antimeron crystals in noncentrosymmetric itinerant magnets on a triangular lattice},
  author = {Hayami, Satoru and Yambe, Ryota},
  journal = {Phys. Rev. B},
  volume = {104},
  issue = {9},
  pages = {094425},
  numpages = {11},
  year = {2021},
  month = {Sep},
  publisher = {American Physical Society},
  doi = {10.1103/PhysRevB.104.094425},
  url = {https://link.aps.org/doi/10.1103/PhysRevB.104.094425}
}

@article{Mohylna25,
  title = {Frustration-driven topological textures on the honeycomb lattice: Antiferromagnetic meron-antimeron and skyrmion crystals emerging from spiral spin liquids},
  author = {Mohylna, M. and G\'omez Albarrac\'{\i}n, F. A. and \ifmmode \check{Z}\else \v{Z}\fi{}ukovi\ifmmode \check{c}\else \v{c}\fi{}, M. and Rosales, H. D.},
  journal = {Phys. Rev. B},
  volume = {111},
  issue = {17},
  pages = {174435},
  numpages = {16},
  year = {2025},
  month = {May},
  publisher = {American Physical Society},
  doi = {10.1103/PhysRevB.111.174435},
  url = {https://link.aps.org/doi/10.1103/PhysRevB.111.174435}
}

@article{Wang26,
doi = {10.1088/1674-1056/ae2672},
url = {https://doi.org/10.1088/1674-1056/ae2672},
year = {2026},
month = {feb},
publisher = {Chinese Physical Society and IOP Publishing Ltd},
volume = {35},
number = {2},
pages = {027502},
author = {Wang, Sidi and Li, Jiyuan and Wang, Yuhao and Xia, Keqi and Meng, Jing and Yu, Bocheng and Hu, Yiqian and Li, Zheng and Zhang, Hui and Luo, Jingzhong and Jiang, Dongmei and Zhan, Qingfeng and Shang, Tian and Xu, Yang},
title = {Hall anomalies in the centrosymmetric triangular-lattice antiferromagnet {{GdGa$_2$}}},
journal = {Chinese Physics B}
}

@article{Baral25,
author = {Baral, Priya Ranjan and Khanh, Nguyen Duy and Gen, Masaki and Sagayama, Hajime and Nakao, Hironori and Arima, Taka-hisa and Ōnuki, Yoshichika and Tokura, Yoshinori and Hirschberger, Max},
title = {Triangular Lattice Magnet {{GdGa$_2$}} with Short-Period Spin Cycloids and Possible Skyrmion Phases},
journal = {Journal of the Physical Society of Japan},
volume = {94},
number = {2},
pages = {024705},
year = {2025},
doi = {10.7566/JPSJ.94.024705}
}

@article{Prychynenko2018,
author = {Prychynenko, Diana and Sitte, Matthias and Litzius, Kai and Kr{\"u}ger, Benjamin and Bourianoff, George and Kl{\"a}ui, Mathias and Sinova, Jairo and Everschor-Sitte, Karin},
doi = {10.1103/PhysRevApplied.9.014034},
journal = {Physical Review Applied},
pages = {14034},
publisher = {American Physical Society},
title = {{Magnetic Skyrmion as a Nonlinear Resistive Element: A Potential Building Block for Reservoir Computing}},
volume = {9},
year = {2018}
}

@article{Song2020a,
author = {Song, Kyung Mee and Jeong, Jae Seung and Pan, Biao and Zhang, Xichao and Xia, Jing and Cha, Sunkyung and Park, Tae Eon and Kim, Kwangsu and Finizio, Simone and Raabe, J{\"{o}}rg and Chang, Joonyeon and Zhou, Yan and Zhao, Weisheng and Kang, Wang and Ju, Hyunsu and Woo, Seonghoon},
doi = {10.1038/s41928-020-0385-0},
journal = {Nature Electronics},
number = {3},
pages = {148--155},
publisher = {Springer US},
title = {{Skyrmion-based artificial synapses for neuromorphic computing}},
volume = {3},
year = {2020}
}

@article{Yokouchi2022,
author = {Yokouchi, Tomoyuki and Sugimoto, Satoshi and Rana, Bivas and Seki, Shinichiro and Ogawa, Naoki and Shiomi, Yuki and Kasai, Shinya and Otani, Yoshichika},
doi = {10.1126/sciadv.abq5652},
journal = {Science Advances},
number = {39},
pages = {eabq5652},
title = {{Pattern recognition with neuromorphic computing using magnetic field–induced dynamics of skyrmions}},
volume = {8},
year = {2022}
}

@article{Vakili2021,
author = {Vakili, Hamed and Zhou, Wei and Ma, Chung T. and Poon, S. J. and Morshed, Md Golam and Sakib, Mohammad Nazmus and Ganguly, Samiran and Stan, Mircea and Hartnett, Timothy Q. and Balachandran, Prasanna and Xu, Jun Wen and Quessab, Yassine and Kent, Andrew D. and Litzius, Kai and Beach, Geoffrey S.D. and Ghosh, Avik W.},
doi = {10.1063/5.0046950},
journal = {Journal of Applied Physics},
number = {7},
publisher = {AIP Publishing LLC},
title = {{Skyrmionics-Computing and memory technologies based on topological excitations in magnets}},
volume = {130},
pages = {070908},
year = {2021}
}

@article{Raju2021,
author = {Raju, M and Petrovi{\'{c}}, A P and Yagil, A and Denisov, K S and Duong, N K and G{\"{o}}bel, B and Auslaender, O M and Mertig, I and Rozhansky, I V and Panagopoulos, C},
doi = {10.1038/s41467-021-22976-6},
journal = {Nature Communications},
number = {2021},
pages = {2758},
title = {{Colossal topological Hall effect at the transition between isolated and lattice-phase interfacial skyrmions}},
volume = {12},
year = {2021}
}

@article{Psaroudaki2023,
author = {Psaroudaki, Christina and Peraticos, Elias and Panagopoulos, Christos},
doi = {10.1063/5.0177864},
journal = {Applied Physics Letters},
pages = {260501},
publisher = {AIP Publishing LLC},
title = {{Skyrmion qubits: Challenges for future quantum computing applications}},
volume = {123},
year = {2023}
}

@article{Tokura2021,
author = {Tokura, Yoshinori and Kanazawa, Naoya},
doi = {10.1021/acs.chemrev.0c00297},
journal = {Chemical Reviews},
pages = {2857},
title = {{Magnetic Skyrmion Materials}},
volume = {121},
year = {2021}
}

@article{Gobel2021,
author = {G{\"{o}}bel, B{\"{o}}rge and Mertig, Ingrid and Tretiakov, Oleg A.},
doi = {10.1016/j.physrep.2020.10.001},
journal = {Physics Reports},
pages = {1--28},
publisher = {Elsevier B.V.},
title = {{Beyond skyrmions: Review and perspectives of alternative magnetic quasiparticles}},
volume = {895},
year = {2021}
}

@article{Wild2017,
author = {Wild, J. and Meier, T. N. G. and P{\"{o}}llath, S. and Kronseder, M. and Bauer, A. and Chacon, A. and Halder, M. and Schowalter, M. and Rosenauer, A. and Zweck, J. and M{\"{u}}ller, J. and Rosch, A. and Pfleiderer, C. and Back, C. H.},
doi = {10.1126/sciadv.1701704},
journal = {Science Advances},
number = {September},
pages = {e1701704},
title = {{Entropy-limited topological protection of skyrmions}},
volume = {3},
year = {2017}
}

@article{Xia2023,
author = {Xia, Jing and Zhang, Xichao and Liu, Xiaoxi and Zhou, Yan and Ezawa, Motohiko},
doi = {10.1103/PhysRevLett.130.106701},
journal = {Physical Review Letters},
number = {10},
pages = {106701},
publisher = {American Physical Society},
title = {{Universal Quantum Computation Based on Nanoscale Skyrmion Helicity Qubits in Frustrated Magnets}},
volume = {130},
year = {2023}
}

@article{Roessler2006,
author = {R{\"{o}}{\ss}ler, U. K. and Bogdanov, A. N. and Pfleiderer, C.},
doi = {10.1038/nature05056},
journal = {Nature},
number = {7104},
pages = {797--801},
title = {{Spontaneous skyrmion ground states in magnetic metals}},
volume = {442},
year = {2006}
}

@article{Lin2016b,
author = {Lin, Shi Zeng and Hayami, Satoru},
doi = {10.1103/PhysRevB.93.064430},
journal = {Physical Review B},
number = {6},
pages = {064430},
title = {{Ginzburg-Landau theory for skyrmions in inversion-symmetric magnets with competing interactions}},
volume = {93},
year = {2016}
}

@article{Arai2026,
author = {Arai, Yuki and Nakayama, Kosuke and Honma, Asuka and Souma, Seigo and Shiga, Daisuke and Kumigashira, Hiroshi and Takahashi, Takashi and Segawa, Kouji and Sato, Takafumi},
doi = {10.1038/s41467-026-71020-y},
journal = {Nature communications},
number = {1},
pages = {3162},
title = {{Origin of multiple skyrmion phases in EuAl4}},
volume = {17},
year = {2026}
}

@article{Bouaziz2022,
author = {Bouaziz, Juba and Mendive-Tapia, Eduardo and Bl{\"{u}}gel, Stefan and Staunton, Julie B.},
doi = {10.1103/PhysRevLett.128.157206},
journal = {Physical Review Letters},
number = {15},
pages = {157206},
publisher = {American Physical Society},
title = {{Fermi-Surface Origin of Skyrmion Lattices in Centrosymmetric Rare-Earth Intermetallics}},
volume = {128},
year = {2022}
}

@article{Kawamura2025,
author = {Kawamura, Hikaru},
doi = {10.1088/1361-648X/adbf5b},
journal = {Journal of Physics Condensed Matter},
number = {18},
title = {{Frustration-induced skyrmion crystals in centrosymmetric magnets}},
volume = {37},
pages = {183004},
year = {2025}
}

@article{Ozawa2017,
author = {Ozawa, Ryo and Hayami, Satoru and Motome, Yukitoshi},
doi = {10.1103/PhysRevLett.118.147205},
journal = {Physical Review Letters},
number = {14},
pages = {147205},
title = {{Zero-Field Skyrmions with a High Topological Number in Itinerant Magnets}},
volume = {118},
year = {2017}
}

@article{Wang2020,
author = {Wang, Zhentao and Batista, Cristian D},
doi = {10.1103/PhysRevB.101.184432},
journal = {Physical Review B},
pages = {184432},
publisher = {American Physical Society},
title = {{Resistivity minimum in diluted metallic magnets}},
volume = {101},
year = {2020}
}

@article{Wang2023f,
author = {Wang, Zhentao and Batista, Cristian D.},
doi = {10.21468/SciPostPhys.15.4.161},
journal = {SciPost Physics},
number = {4},
pages = {161},
title = {{Skyrmion crystals in the triangular Kondo lattice model}},
volume = {15},
year = {2023}
}

@article{Petrovic2025,
author = {Petrovi{\'{c}}, Alexander P and Psaroudaki, Christina and Fischer, Peter and Garst, Markus and Panagopoulos, Christos},
doi = {10.1103/4jwn-d6dg},
journal = {Reviews of Modern Physics},
number = {September},
pages = {031001},
title = {{Colloquium: Quantum properties and functionalities of magnetic skyrmions}},
volume = {97},
year = {2025}
}

@article{Matsumura2024a,
author = {Matsumura, Takeshi and Kurauchi, Kenshin and Tsukagoshi, Mitsuru and Higa, Nonoka and Nakao, Hironori and Kakihana, Masashi and Hedo, Masato and Nakama, Takao and Ōnuki, Yoshichika},
doi = {10.7566/JPSJ.93.074705},
journal = {Journal of the Physical Society of Japan},
number = {7},
pages = {074705},
title = {{Helicity Unification by Triangular Skyrmion Lattice Formation in the Noncentrosymmetric Tetragonal Magnet EuNiGe3}},
volume = {93},
year = {2024}
}

@article{Matsumura2024,
author = {Matsumura, Takeshi and Tabata, Chihiro and Kaneko, Koji and Nakao, Hironori and Kakihana, Masashi and Hedo, Masato and Nakama, Takao and Ōnuki, Yoshichika},
doi = {10.1103/PhysRevB.109.174437},
journal = {Physical Review B},
number = {17},
pages = {174437},
title = {{Single helicity of the triple- q triangular skyrmion lattice state in the cubic chiral helimagnet EuPtSi}},
volume = {109},
year = {2024}
}

@article{Yao2020,
author = {Yao, Xiaoyan and Chen, Jun and Dong, Shuai},
doi = {10.1088/1367-2630/aba1b3},
journal = {New Journal of Physics},
number = {8},
pages = {083032},
publisher = {IOP Publishing},
title = {{Controlling the helicity of magnetic skyrmions by electrical field in frustrated magnets}},
volume = {22},
year = {2020}
}

@article{Marrows2021,
author = {Marrows, C H and Zeissler, K},
doi = {10.1063/5.0072735},
journal = {Appl. Phys. Lett.},
pages = {250502},
publisher = {AIP Publishing LLC},
title = {{Perspective on skyrmion spintronics}},
volume = {119},
year = {2021}
}

@article{Hals2016,
author = {Hals, Kjetil M.D. and Schecter, Michael and Rudner, Mark S.},
doi = {10.1103/PhysRevLett.117.017001},
journal = {Physical Review Letters},
number = {1},
title = {{Composite Topological Excitations in Ferromagnet-Superconductor Heterostructures}},
pages = {017001},
volume = {117},
year = {2016}
}

@article{Baumard2019,
author = {Baumard, J. and Cayssol, J. and Bergeret, F. S. and Buzdin, A.},
doi = {10.1103/PhysRevB.99.014511},
journal = {Physical Review B},
pages = {014511},
publisher = {American Physical Society},
title = {{Generation of a superconducting vortex via N\'eel skyrmions}},
volume = {99},
year = {2019}
}

@article{Wu2024,
author = {Wu, Kai and Zhao, Yuelei and Hao, Hongyuan and Yang, Sheng and Li, Shuang and Liu, Qingfang and Zhang, Senfu and Zhang, Xixiang and {\AA}kerman, Johan},
doi = {10.1038/s41467-024-54851-5},
journal = {Nature Communications},
pages = {10463},
title = {{Topological transformation of synthetic ferromagnetic skyrmions : thermal assisted switching of helicity by spin-orbit torque}},
volume = {15},
year = {2024}
}

@article{Dai2023,
author = {Dai, Bingqian and Wu, Di and Razavi, Seyed Armin and Xu, Shijie and He, Haoran and Shu, Qingyuan and Jackson, Malcolm and Mahfouzi, Farzad and Huang, Hanshen and Pan, Quanjun and Cheng, Yang and Qu, Tao and Wang, Tianyi and Tai, Lixuan and Wong, Kin and Kioussis, Nicholas and Wang, Kang L},
doi = {10.1126/sciadv.ade6836},
journal = {Science Advances},
number = {February},
pages = {eade6836},
title = {{Electric field manipulation of spin chirality and skyrmion dynamic}},
volume = {9},
year = {2023}
}

@article{Lv2022,
author = {Lv, Xiaowei and Pei, Ke and Yang, Chendi and Qin, Gang and Liu, Min and Zhang, Jincang and Che, Renchao},
doi = {10.1021/acsnano.2c08844},
journal = {ACS Nano},
pages = {19319},
title = {{Controllable Topological Magnetic Transformations in the Thickness-Tunable van der Waals Ferromagnet Fe$_5$GeTe$_2$}},
volume = {16},
year = {2022}
}

@article{Li2023g,
author = {Li, Long and Song, Dongsheng and Wang, Weiwei and Zheng, Fengshan and Kov{\'{a}}cs, Andr{\'{a}}s and Tian, Mingliang and Dunin-borkowski, Rafal E and Du, Haifeng},
doi = {10.1002/adma.202209798},
journal = {Advanced Materials},
pages = {2209798},
title = {{Transformation from Magnetic Soliton to Skyrmion in a Monoaxial Chiral Magnet}},
volume = {35},
year = {2023}
}

@article{Verba2020,
author = {Verba, R. V. and Navas, D. and Bunyaev, S. A. and Hierro-Rodriguez, A. and Guslienko, K. Y. and Ivanov, B. A. and Kakazei, G. N.},
doi = {10.1103/PhysRevB.101.064429},
journal = {Physical Review B},
number = {6},
pages = {064429},
publisher = {American Physical Society},
title = {{Helicity of magnetic vortices and skyrmions in soft ferromagnetic nanodots and films biased by stray radial fields}},
volume = {101},
year = {2020}
}

@article{Kong2024,
author = {Kong, Lingyao and Tang, Jin and Wu, Yaodong and Wang, Weiwei and Jiang, Jialiang and Wang, Yihao and Li, Junbo and Xiong, Yimin and Wang, Shouguo and Tian, Mingliang and Du, Haifeng},
doi = {10.1103/PhysRevB.109.014401},
journal = {Physical Review B},
number = {1},
pages = {014401},
publisher = {American Physical Society},
title = {{Diverse helicities of dipolar skyrmions}},
volume = {109},
year = {2024}
}

@article{Nayak2017,
author = {Nayak, Ajaya K. and Kumar, Vivek and Ma, Tianping and Werner, Peter and Pippel, Eckhard and Sahoo, Roshnee and Damay, Fran{\c{c}}oise and R{\"{o}}{\ss}ler, Ulrich K. and Felser, Claudia and Parkin, Stuart S.P.},
doi = {10.1038/nature23466},
journal = {Nature},
number = {7669},
pages = {561--566},
title = {{Magnetic antiskyrmions above room temperature in tetragonal Heusler materials}},
volume = {548},
year = {2017}
}
\end{document}